\RequirePackage{fix-cm}
\documentclass[amstex,10pt,reqno]{amsart}
\usepackage{amsmath,amsfonts,amssymb,enumerate,hyperref,multicol,color,graphicx,epstopdf}
\begin{document}
\vspace{-1cm}
\title{GFEM Study of magnetohydrodynamics thermo-diffusive effect on nanofluid flow over power-law stretching sheet along with regression analysis}
\maketitle
\begin{center}
{\bf Rangoli Goyal$^1$ and Rama Bhargava$^2$}
\vskip0.2in
$^{1,2}$
Department of Mathematics, Indian Institute of Technology Roorkee, India\\

\vskip0.2in

 $^1$rangoligoyal@gmail.com
$^2$rbharfma@iitr.ac.in
\end{center}

\begin{abstract}
The present paper uses the Galerkin Finite Element Method to numerically study the triple diffusive boundary layer flow of homogenous nanofluid over power-law stretching sheet with the effect of external magnetic field. The fluid is composed of nanoparticles along with dissolved solutal particles in the base fluid. The chief mechanisms responsible for enhancement of convective transport phenomenon in nanofluids - Brownian Motion, Diffusiophoresis and Thermophoresis have been considered. The simulations performed in this study are based on the boundary layer approach. Recently proposed heat flux and nanoparticle mass flux boundary conditions have been imposed. Heat transfer, solutal mass transfer and nanoparticle mass transfer are investigated for different values of controlling parameters i.e. Brownian-motion parameter, Thermophoresis parameter, magnetic influence parameter and stretching parameter. Multiple regression analysis has been performed to verify the relationship among transfer rate parameters and controlling parameters. The present study finds application in insulation of wires, manufacture of tetra packs, production of glass fibres, fabrication of various polymer and plastic products, rubber sheets etc. where the quality merit of desired product depends on the rate of stretching, external magnetic field and composition of materials used.
\vspace{0.2cm}
\begin{flushleft}
\textbf{Keywords:} Brownian Motion; Thermophoresis; FEM; Magnetic field; Multiple Regression Estimate.
\end{flushleft}
\end{abstract}

\onecolumn
\begin{center}
Nomenclature
\end{center}
\begin{tabular}{l l}
$m$ & nonlinear stretching parameter\\
$u$  & velocity in x-direction \\
$v$  & velocity in y-direction  \\
$B$ & Magnetic field\\
$Bi$ & Biot number\\
$D_s$  & solutal diffusivity \\
$D_b$  & Brownian diffusion coefficient\\
$D_t$ & Thermophoresis coefficient \\
$D_{TC}$ & Dufour diffusivity\\
$D_{CT}$ & Soret diffusivity\\
$Ln$ & nanofluid Lewis number\\
$Le$ & regular Lewis number\\
$Ld$ & Dufour solutal Lewis number\\
$M$ & nondimensional magnetic parameter\\
$Nd$ & modified Dufour parameter\\
$Nb$ & Brownian motion parameter\\
$Nur$ & Modified Nusselt number\\
$Nt$ & Thermophoresis parameter\\
$Pr$ & Prandtl number\\
$Shr$ & Modified Sherwood number\\
\end{tabular}
\begin{tabular}{l l}
$Shrn$ & Modified nanofluid Sherwood number\\
$T$  & temperature \\
$\phi_{s}$  & solutal concentration \\
$\phi_{np}$  & nanoparticle concentration \\
$\alpha_{m}$  & thermal diffusivity  \\
$\tau$  & ratio of nanoparticle heat capacity to\\
  & fluid heat capacity \\
$\rho$  & effective fluid density \\
$\mu$  & effective dynamic viscosity \\
$k$  & thermal conductivity \\
$\rho c_{p}$  & heat capacity\\
$\theta$ & non-dimensional temperature\\
$\gamma$ & non-dimensional solutal concentration\\
$f$ & non-dimensional nanoparticle concentration\\
$np$ & solid particle\\
$bf$ & base fluid\\
$nf$ & nanofluid\\
$w$ & boundary layer of sheet\\
$\infty$ & outer layer of boundary layer flow
\end{tabular}

\section{Introduction}
\label{intro}
The analysis of heat and mass transfer along boundary layer helps in better designing of devices involving ultra high cooling applications. Water, toulene, ethylene glycol, or oil are the most commonly used fluids for heat transfer. The characteristic drawback of inherently poor thermal conductivity renders them inefficient for the choice of competent heat transfer fluids. It has been observed that addition of metallic solid particles helps in enhancing the thermal conductivity. The suspension of micro sized particles in fluid comes with a disadvantage of clogging, abrasion and sedimentation. Choi and Eastman \cite{Ref291} observed that nanometre sized particles with increased surface area per unit volume provide a better alternative to micro size particles and concluded that their suspension greatly enhances the thermal properties of the fluid.  Choi et al. \cite{Ref281} proposed the term 'nanofluid' for this new class of fluids containing suspended nanometre sized particles in the base fluid. Buongiorno \cite{Ref4} performed a detailed study of the mechanisms responsible for convective transport in nanofluids, concluding that seven slip mechanisms are responsible for intensifying the heat transfer. The above study considered the two component nanofluid model, taking into account the percentage of volume of nanoparticle diffused in base fluid.Kuznetsov and Nield \cite{Ref13} extended this model for convective boundary conditions considering that nanoparticle concentration is passively controlled.\\
The important engineering applications concerning the fluid flow on top of a non-linear stretching sheet are found in the field of metallurgical and chemical engineering, particularly in the process that involve drawing a continuous strip or filament through an inert fluid and reconstituting it as finished component. The desired strength and stiffness of components is achieved by controlling the rate of stretching and time rate cooling of the fluid along the sheet.
The first study of boundary layer flow of fluid on top of a stretching sheet was done by Sakiadis \cite{Ref23}. He studied the flow for the linear stretching sheet moving with a constant velocity.  Crane \cite{Ref30} made an extension to the problem by considering a sheet moving with velocity which is in proportion to distance from the slit. The above studies drew considerable attention from researchers and paved way for studies involving study of flow and heat transfer of fluids on top of a stretching surface \cite{Ref2}, \cite{Ref13}, \cite{Ref14}. All the above mentioned studies were conducted by considering either a linearly stretching sheet problem or a constant value for the velocity wall. \\
Depending upon the characteristics of the final product, linear stretching might not be a feasible option. The viscous boundary layer flow atop a non-linear stretching sheet was investigated by Vajravelu \cite{Ref25}. He used the fourth order Runge-Kutta integration to solve the system of equation. Cortell \cite{Ref6} presented an analysis for fluid flow atop a permeable surface which is moving with velocity $u_{w} (x) $ = $x^{1/3}$. He considered constant-surface-temperature (CST) and prescribed-surface-temperature (PST) boundary condition on the sheet. Kumaran and Ramaniah \cite{Ref12} studied flow over a stretching sheet which is moving with velocity having a quadratic expression subject to linear mass flux boundary condition. Kechil and Hashim \cite{Ref10} derived the series solution for MHD flow atop a non linear stretching sheet in existence of chemical reaction using Adomian decomposition method. Hamad et. al. \cite{Ref9} derived the similarity solution for nanofluid flow atop a nonlinear stretching sheet considering the two phase model foe nanofluid.  Ziabakhsh et al. \cite{Ref31} provided the analytical solution for diffusion of chemically reactive species  over a non-linear stretching sheet embedded in porous medium. Narayana and Sibanda \cite{Ref16} investigated a parametric study of laminar flow atop an unsteady stretching sheet using the EMT (effective medium theory) model of nanofluids. Goyal and Bhargava \cite{Ref8} investigated the diffusive-thermo and thermo-diffusive effects of nanofluid flow atop a power law stretching sheet.\\
In various real life industrial processes, the system has a number of components whose concentrations vary with respect to time. As the system tends to achieve homogeneity, the components from area of higher concentration move to area of lower concentration such that the concentration difference is minimized. This results in an energy flux being created by both temperature gradient as well as composition gradient. Thermo-diffusion effect or Soret effect is the mass flux created because of temperature gradient. Diffusio- thermo effect or Dufour effect is the energy flux created by the composition gradient. Pop and Ingham \cite{Ref121} presented a detailed review about the boundary layer heat transfer research identifying the possible difficulties and future requirements to be. Khan et al. \cite{Ref131} studied the effect of thermo slip and hydrodynamic slip boundary conditions on free convective nanofluid fluid along a vertical plate.  \\

The heat and mass transfer properties of the nanofluid can be controlled by regulating the amount and type of nanoparticles and base fluid being used. Magnetohydrodynamics (MHD) effects also play and influential role in controlling the rate of cooling as well as segregation of molten metals from various non$-$metallic impurities. As the name suggests, MHD effects refer to the movement of object (\textit{dynamics}) by magnetic force (\textit{magneto}) inside water or fluid (\textit{hydro}).The presence of external magentic field gives rise to Lorentz drag force which acts on the fluid, thereby potentially altering the fluid flow characteristics such as velocity, temperature and concentration. The increasing need of innovation for better design of products in various metallurgical and polymer extrusion processes has led to an interest in study of MHD flow over a moving plate \cite{Ref17}, \cite{Ref18}. This type of flow finds application in glass fiber drawing and paper production, chilling of metallic sheetsand electronic chips, heat-treated materials traveling in between a feed roll and a wind-up roll,  crystal growing etc.\\
Chiam \cite{Ref5} investigated the boundary layer flow over a plate moving with power-law velocity under the influence of external transverse magnetic field, deriving an explicit expression for skin friction coefficient followed by a numerical solution using a shooting method. The influence of Soret and Dufour diffusion on laminar MHD mixed convection boundary layer flow along a vertical stretching surface was studied by Beg et al. \cite{Ref3}. Dandapat et. al. \cite{Ref7} considered the stretching velocity and temperature distribution in general functional forms while studying the flow of a fine liquid film along a horizontal stretching surface under transverse magnetic field; concluding that magnetic field resists thinning of the film. Martin et. al. \cite{Ref15} studied the mixed convection magneto hydrodynamic nanofluid flow under the effect of viscous dissipation and variable external magnetic field along a sheet being stretched with power law velocity. Recently, Awad et al. \cite{Ref1} investigated numerically the thermodiffusion effects on magneto-nanofluid fluid over a stretching sheet.  \\
The present work extends the study of Goyal and Bhargava \cite{Ref8} with modification of Buongiorno's nanofluid model for the case of binary nanofluid. The nanofluid contains solute particles, thereby enabling cross diffusion to occur. Variable magnetic field is considered int he problem. This problem studies the triple diffusion (Brownian motion diffusion, Thermophoresis and Diffusiophoresis). Finite Element Method is used to obtain numerical solution for the problem and regression analysis is conducted to verify the relationship among parameters. The effects of restraining parameters on heat and mass transfer rate has been demonstrated both graphically and in tabular form.

\section{Problem Description $\&$ Mathematical Development}
A steady state two-dimensional natural convection boundary layer flow of $Al_{2}O_{3}$-water nanofluid along a nonlinear stretching sheet is considered. The sheet is saturated in a binary fluid medium which has dissolved solute particles and dispersed nanoparticles. The sheet coincides with the plane $y=0$. It is being stretched in $x$-direction with velocity $U=u_{w}=ax^{m}$. The external forces (gravity) and pressure gradient are neglected. The temperature of the fluid $T_{bf}$ is assumed to be much larger than the ambient temperature $T_{\infty}$.  \\
For incompressible fluid flow, with boundary layer approximation, the mathematical equations as derived by  Buongiorno \cite{Ref4} and Khan et. al. \cite{Ref11} illustrating conservation of mass, momentum, temperature, solutal concentration and nanoparticle concentration respectively is as follows
\begin{equation}
\frac{\partial u}{\partial x} + \frac{\partial v}{\partial y} = 0
\end{equation}
\begin{equation}
u \frac{\partial u}{\partial x} +  v \frac{\partial u}{\partial y} + \frac{\sigma}{\rho_{nf}} B^{2} u = \nu  \frac{\partial^{2} u}{
\partial y^{2}}
\end{equation}
\begin{equation}
u \frac{\partial T}{\partial x} +  v \frac{\partial T}{\partial y} = \alpha_{m} \frac{\partial^{2} T}{\partial y^{2}} + \tau \bigg( D_{B} \frac{\partial \widehat{\phi}}{\partial y} \frac{\partial T}{\partial y} +  \frac{D_{T}}{T_{\infty}} \bigg( \frac{\partial T}{\partial y} \bigg)^{2} \bigg) + D_{TC} \frac{\partial^{2} \phi_{s}}{\partial y^{2}}
\end{equation}
\begin{equation}
u \frac{\partial \phi_{s}}{\partial x} +  v \frac{\partial \phi_{s}}{\partial y} = D_{S} \frac{\partial^{2} \phi_{s}}{\partial y^{2}} + D_{CT} \frac{\partial^{2} T}{\partial y^{2}}
\end{equation}
\begin{equation}
u  \frac{\partial \phi_{np}}{\partial y} + v  \frac{\partial \phi_{np}}{\partial y} = D_{B} \frac{\partial^{2} \phi_{np}}{\partial y^{2}} + \frac{D_{T}}{T_{\infty}}\frac{\partial^{2} T}{\partial y^{2}}
\end{equation}
The boundary conditions are as follows
\begin{equation}
u_{w}=ax^{m}, \;  v=0, \;  \phi_{s} = \phi_{sw}, \;  -k \frac{\partial T}{\partial y} = h_{bf} (T_{bf} - T), \; D_{B}\frac{\partial \phi_{np}}{\partial y} + \frac{D_{T}}{T_{\infty}} \frac{\partial T}{\partial y} \;at  \; y=0
\end{equation}
\begin{equation}
u\rightarrow 0, \;  T \rightarrow T_{\infty}, \; \phi_{s}\rightarrow \phi_{s \infty} \;  \textsl{and}  \; \phi_{np} = \phi_{np \infty}  \;    \textsl{when}  \;when  \; y \rightarrow \infty
\end{equation}

The last term of left hand side of equation (2) represents the magnetic flux. The functional expression for magnetic field is given by $B(x)$ = $B_{0} x^{\frac{m-1}{2}}$.
The new set of dimensionless parameters defined to transform Equations (1-5) and (6), (7) into a set of non dimensional equations are:
\begin{center}
$\eta = y \sqrt{\frac{a(m+1)}{2 \nu}} x^{\frac{m-1}{2}}, u = ax^{m}s'(\eta), $ \\$ v=-\sqrt{\frac{a(m+1)}{2 }} x^{\frac{m-1}{2}} \times  \bigg( s(  \eta) + \frac{m-1}{m+1}  \eta s'(\eta)  \bigg),$
\end{center}
\begin{equation}
 \theta(\eta) = \frac{T - T_{\infty}}{T_{w}- T_{\infty}}, \gamma (\eta) = \frac{\phi_{s} - \phi_{s \infty}}{\phi_{sw}- \phi_{s \infty}}, f ( \eta) = \frac{\phi_{np} - \phi_{np \infty}}{\phi_{np w}-  \phi_{np \infty}}
\end{equation}
The continuity equation is compulsorily satisfied. The transformed momentum, energy and concentration equations are thus formulated as follows
\begin{equation}
s'''+ss'' - \frac{2m}{m+1}s'^{2}- Ms'=0
\end{equation}
\begin{equation}
\frac{1}{Pr} \theta''+ s \theta' + Nb \theta'f' + Nt \theta'^{2}+Nd \gamma ''= 0
\end{equation}
\begin{equation}
\gamma''+ Le s\gamma'+Ld \theta '' = 0
\end{equation}
\begin{equation}
f'' + Ln s f' + \frac{Nt}{Nb} \theta '' = 0
\end{equation}
The corresponding boundary conditions are transformed to,
\begin{equation}
\begin{split}
s(0)=0, s'(0)=1,  \theta^{'} = -Bi ( 1 - \theta (0)), Nbf^{'}(0) + Nt \theta^{'}(0) =0,  f(0)=1,  \text{ at } \eta = 0\\
s'\rightarrow 0, \theta \rightarrow 0,\gamma \rightarrow 0, f \rightarrow 0, \text{ as } \eta \rightarrow \infty
\end{split}
\end{equation}
where primes denote differentiation with respect to $\eta$. The value of $Bi$ is assumed to be $0.01$ throughout the paper unless mentioned otherwise. The parameters appearing in equations are defined as follows:
\begin{equation}
\begin{split}
Pr = \frac{\nu}{\alpha_{m}}, Ln = \frac{\nu}{D_{B}}
, Le = \frac{\nu}{D_{S}}, Ld = \frac{D_{CT}(T_{w}-T_{\infty})}{D_{S}(\phi_{sw}- \phi_{s \infty})},
Nd= \frac{D_{TC}(\phi_{sw}- \phi_{s \infty})}{\nu(T_{w}-T_{\infty})} \\
M=\frac{2 \sigma B_{o}^{2}}{\rho_{bf} a (m+1)},
Nb=\frac{(\rho c)_{p} D_{B}(\phi_{npw}- \phi_{np \infty})}{(\rho c)_{bf} \nu}, Nt=\frac{(\rho c)_{p} D_{T}(T_{w}-T_{\infty})}{(\rho c)_{bf} T_{\infty} \nu}
\end{split}
\end{equation}
The variables of significant practical interest in this study are the Nusselt number, the Sherwood number and the nanofluid Sherwood number, defined as
\begin{equation}
Nu_{x} = \frac{xq_{w}}{k(T_{w} - T_{\infty})}, Sh_{x} = \frac{xq_{m}}{D_{S}(C_{w} - C_{\infty})}, Sh_{x,n} = \frac{xq_{np}}{D_{B}(\phi_{w} - \phi_{\infty})}
\end{equation}
where $q_{w}$, $q_{m}$ and $q_{np}$ represent the heat flux, regular mass flux and nano mass flux at the sheet surface respectively. They are given by the following expressions
\begin{equation}
q_{w} = -k \bigg( \frac{\partial T} {\partial y} \bigg)_{y=0}, q_{m} = -D_{S} \bigg( \frac{\partial \phi_{s}} {\partial y} \bigg)_{y=0}, q_{np} = -D_{B} \bigg( \frac{\partial \phi_{np}} {\partial y} \bigg)_{y=0}
\end{equation}
Reduced expression for Nusselt number, Sherwood number and nanofluid Sherwood number is expressed as
\begin{equation}
\begin{split}
Nur = -\theta^{'} (0) = \frac{Nu_{x}}{\sqrt{\frac{m+1}{2} Re_{x}}}, Shr = -\gamma^{'} (0) = \frac{Sh_{x}}{\sqrt{\frac{m+1}{2} Re_{x}}}, \\
Shrn = -f^{'} (0) = \frac{Sh_{x,n}}{\sqrt{\frac{m+1}{2} Re_{x}}}
\end{split}
\end{equation}
$Re_{x}$ is the local Reynolds number whose value is based on the velocity $u_{w}(x)$ with which the sheet is being stretched. The reduced Nusselt number proportional to $-\theta^{'} (0)$ gauges the effect of heat transfer in the system.
Similarly, reduced Sherwood number proportional to $-\gamma^{'} (0)$ and reduced nanofluid Sherwood number proportional to $-f^{'} (0)$ are used to gauge the effect of solutal mass transfer and nanoparticle mass transfer respectively.

\section{Numerical Implementation and Post Processing}
\textbf{Finite Element Method}\\
The Galerkin Finite Element Method (GFEM) is a numerical method highly efficient in providing approximate solutions to system of partial differential equations, which form the governing equations for various practical engineering problems arising in different areas such as, fluid mechanics \cite{Ref22}, rigid body dynamics \cite{Ref33}, chemical processing \cite{Ref32},  solid mechanics \cite{Ref34}  etc. It is one of the most powerful method in context of its implementation to real-world problems involving complex geometry and/or complicated boundary conditions. As the name suggests, the basic concept lies in dividing the whole domain into smaller elements of finite dimensions. \\
The given domain is broken down into a number of subdomains, and over every subdomain, the approximate solution of governing equation is estimated by any of the conventional variational methods. Finding approximate solution on the subdomains helps in representing a complex function as an assemblage of simple polynomials. The essential fact to be taken care of is to presume the piecewise continuity of the function for acquiring the solution. We have a set of partial differential equations, given in (9-12), with (13) representing the boundary conditions. To compute the solution of these equations, presume that
\begin{equation}
s'=h
\end{equation}
After substituting this, the system of equations (9-12) reduces to
\begin{equation}
h''+sh' - \frac{2m}{m+1}h^{2}- Mh=0
\end{equation}
\begin{equation}
\frac{1}{Pr} \theta''+ s \theta' + Nb \theta'f' + Nt \theta'^{2}+Nd \gamma ''= 0
\end{equation}
\begin{equation}
\gamma''+ Le s\gamma'+Ld \theta '' = 0
\end{equation}
\begin{equation}
f'' + Ln s f' + \frac{Nt}{Nb} \theta '' = 0
\end{equation}
and the corresponding boundary conditions are as follows:
\begin{equation}
\begin{split}
 h(0)=1,  \theta^{'} = -Bi ( 1 - \theta (0)), Nbf^{'}(0) + Nt \theta^{'}(0) =0,  f(0)=1,  \text{ at } \eta = 0\\
s'\rightarrow 0, \theta \rightarrow 0,\gamma \rightarrow 0, f \rightarrow 0, \text{ as } \eta \rightarrow \infty
\end{split}
\end{equation}
The weighted residual formulation of the given differential equations over the typical linear element denoted by $\Omega_{e}$ having coordinates $( \eta_{e}, \eta_{e+1})$, is given by
\begin{equation}
\begin{split}
\int_{\eta_{e}}^{\eta_{e+1}} W_{1} \lbrace s' - h \rbrace d\eta = 0\\ \int_{\eta_{e}}^{\eta_{e+1}} W_{2}   \bigg\lbrace h''+sh' - \frac{2m}{m+1}h^{2}- Mh  \bigg\rbrace d\eta = 0\\
\int_{\eta_{e}}^{\eta_{e+1}} W_{3} \bigg\lbrace \frac{1}{Pr} \theta''+ s \theta' + Nb \theta'f' + Nt \theta'^{2}+Nd \gamma ''  \bigg\rbrace d\eta = 0\\
\int_{\eta_{e}}^{\eta_{e+1}} W_{4} \bigg\lbrace \gamma''+ Le s\gamma'+Ld \theta ''  \bigg\rbrace d\eta = 0\\
\int_{\eta_{e}}^{\eta_{e+1}} W_{5} \bigg\lbrace f'' + Ln s f' + \frac{Nt}{Nb} \theta ''  \bigg\rbrace d\eta = 0
\end{split}
\end{equation}
where $W_{1}, W_{2}, W_{3}, W_{4} $ and $W_{5}$ are randomly chosen test functions. They can be viewed as variants of $s$, $h$, $\theta$, $\gamma$ and $f$ respectively. \\
A typical dependent variable $\Theta$ of the form $\Theta$ = $\sum_{i=1}^{3} \Theta_{i} \psi_{i}$ is assumed where $\Theta$ stands for either $s$, $h$, $\theta$, $\gamma$ and $f$ with $W_{1}= W_{2}=  W_{3}= W_{4} = W_{5} = \psi_{j}$, $(j=1,2,3,4,5)$.\\
In our calculations, the shape functions for a typical element are taken as quadratic element. The finite element model thus formulated is as follows
\[
\left[ \begin {array}{ccccc}
[K^{11}] & [K^{12}] & [K^{13}] & [K^{14}] & [K^{15}]\\
\left[ K^{21} \right]  & [K^{22}] & [K^{23}] & [K^{24}] & [K^{25}]\\
\left[ K^{31} \right]  & [K^{32}] & [K^{33}] & [K^{34}] & [K^{35}]\\
\left[ K^{41} \right]  & [K^{42}] & [K^{43}] & [K^{44}] & [K^{45}]\\
\left[ K^{51} \right]  & [K^{52}] & [K^{53}] & [K^{54}] & [K^{55}]\\
\end{array} \right]
\left[ \begin {array}{c}
s\\
h\\
\theta\\
\gamma\\
f\\
\end{array} \right]
=
\left[ \begin {array}{c}
\lbrace b^{1} \rbrace\\
\lbrace b^{2} \rbrace\\
\lbrace b^{3} \rbrace\\
\lbrace b^{4} \rbrace\\
\lbrace b^{5} \rbrace\\
\end{array} \right]
\]
where $[ K^{mn} ]$, $(m,n) = {1,2,3,4,5}$ are determined as:
\begin{equation}
\begin{split}
K_{ij}^{11} = \int_{\eta_{e}}^{\eta_{e+1}} \psi_{i} \frac{\partial \psi_{j} }{\partial \eta} d \eta, K_{ij}^{12} = -\int_{\eta_{e}}^{\eta_{e+1}} \psi_{i}\psi_{j} d \eta, K_{ij}^{13} = K_{ij}^{14} = K_{ij}^{15} = 0\\
K_{ij}^{21} = K_{ij}^{23} =K_{ij}^{24} = K_{ij}^{25} = 0\\
K_{ij}^{22} = -\int_{\eta_{e}}^{\eta_{e+1}} \frac{\partial \psi_{i} }{\partial \eta} \frac{\partial \psi_{j} }{\partial \eta} d \eta + \int_{\eta_{e}}^{\eta_{e+1}}  \psi_{i} \overline{s} \frac{\partial \psi_{j} }{\partial \eta} d \eta - \frac{2m}{m+1} \int_{\eta_{e}}^{\eta_{e+1}}  \psi_{i} \overline{h}  \psi_{j}  d \eta - M \int_{\eta_{e}}^{\eta_{e+1}} \psi_{i}\psi_{j} d \eta\\
K_{ij}^{31} = K_{ij}^{32} = K_{ij}^{35} = 0, K_{ij}^{34} = -Nd \int_{\eta_{e}}^{\eta_{e+1}} \frac{\partial \psi_{i} }{\partial \eta} \frac{\partial \psi_{j} }{\partial \eta} d \eta\\
K_{ij}^{33} = - \frac{1}{Pr} \int_{\eta_{e}}^{\eta_{e+1}} \frac{\partial \psi_{i} }{\partial \eta} \frac{\partial \psi_{j} }{\partial \eta} d \eta + \int_{\eta_{e}}^{\eta_{e+1}} \psi_{i} \overline{s} \frac{\partial \psi_{j} }{\partial \eta} d \eta + Nb \int_{\eta_{e}}^{\eta_{e+1}} \psi_{i} \overline{f'} \frac{\partial \psi_{j} }{\partial \eta} d \eta +  Nt \int_{\eta_{e}}^{\eta_{e+1}} \psi_{i} \overline{\theta'} \frac{\partial \psi_{j} }{\partial \eta} d \eta\\
K_{ij}^{41} = K_{ij}^{42} = K_{ij}^{45} = 0, K_{ij}^{43} = -Ld \int_{\eta_{e}}^{\eta_{e+1}} \frac{\partial \psi_{i} }{\partial \eta} \frac{\partial \psi_{j} }{\partial \eta} d \eta,
K_{ij}^{44} = - \int_{\eta_{e}}^{\eta_{e+1}} \frac{\partial \psi_{i} }{\partial \eta} \frac{\partial \psi_{j} }{\partial \eta} d \eta + Le \int_{\eta_{e}}^{\eta_{e+1}} \psi_{i} \overline{s} \frac{\partial \psi_{j} }{\partial \eta} d \eta \\
K_{ij}^{51} = K_{ij}^{52} = K_{ij}^{54} = 0,
K_{ij}^{53} = - \frac{Nt}{Nb} \int_{\eta_{e}}^{\eta_{e+1}} \frac{\partial \psi_{i} }{\partial \eta} \frac{\partial \psi_{j} }{\partial \eta} d \eta, K_{ij}^{55} = - \int_{\eta_{e}}^{\eta_{e+1}} \frac{\partial \psi_{i} }{\partial \eta} \frac{\partial \psi_{j} }{\partial \eta} d \eta+ Ln \int_{\eta_{e}}^{\eta_{e+1}} \psi_{i} \overline{s} \frac{\partial \psi_{j} }{\partial \eta} d \eta
\end{split}
\end{equation}
where
\begin{equation}
\begin{split}
\overline{s} = \sum_{i=1}^{3} \overline{s_{i}} \psi_{i}, \overline{s'} = \sum_{i=1}^{3} \overline{s_{i}} \frac{\partial \psi_{i} }{\partial \eta}, \overline{h} = \sum_{i=1}^{3} \overline{h_{i}} \psi_{i}, \overline{h'} = \sum_{i=1}^{3} \overline{h_{i}} \frac{\partial \psi_{i} }{\partial \eta}, \overline{\theta'} = \sum_{i=1}^{3} \overline{\theta_{i}} \frac{\partial \psi_{i} }{\partial \eta},\\
\overline{\gamma} = \sum_{i=1}^{3} \overline{\gamma_{i}} \psi_{i}, \overline{\gamma'} = \sum_{i=1}^{3} \overline{\gamma_{i}} \frac{\partial \psi_{i}}{\partial \eta},  \overline{f'} = \sum_{i=1}^{3} \overline{f_{i}} \frac{\partial \psi_{i} }{\partial \eta}
\end{split}
\end{equation}
Relating the element nodes to global nodes, an interelement continuity equation is identified for primary variables. The global stiffness matrix is formulated. The Dirichlet boundary conditions and Neumann boundary conditions  are imposed. To ensure that the solution is grid independent, an extensive grid independence test is conducted. It helps in suitably guessing the value of $\eta_{\infty}$ for given boundary value problem. The series of values for $|- \theta '(0)|, |- \gamma' (0)|$ and $| - f'(0)|$ with different values of $\eta_{\infty}$ and step sizes $h$ (from 0.2 to 0.0001) are computed so as to determine the value of $\eta_{\infty}$ for which results are independent of length of domain (Table 1). The domain area for integration $\eta$ is considered from 0 to $\eta_{\infty}$ = 10. $\eta_{\infty}$ corresponds to $\eta \rightarrow \infty$ which lies well outside the boundary layer. From Table 1, observing the sensitivity of solution to grid compactness, it can be noted that for same domain the accuracy is not impacted even if size of elements is decreased or number of elements is increased.  Increasing number of elements or decreasing the size of elements only increases compilation time and does not enhance the accuracy of solutions.\\
The entire flow domain is divided into a set of 5000 quadratic elements of equal length. The quadratic element is three noded and hence, total number of nodes in domain are 10,001. A system of 50,005 nonlinear equations is obtained. An iterative scheme must be used to solve the system of equations and hence, in the present study, Gauss elimination method is employed. Accuracy of $0.5 \times 10^{-4}$ is maintained. Gaussian quadrature method is used to solve the integrations. The relative difference between the current and present iterations is used as the convergence criterion. MATLAB is used to execute the code of the algorithm.

\section{Multiple Regression Estimation (MRE)}
Multiple regression analysis is a statistical tool used for estimating relationship among variables. It allows explicit control of many factors which are simultaneously affecting the dependent variable. Multiple regression estimations, of the modified Nusselt number $Nur_{mre} $ and modified Sherwood number $Shr_{mre}$ have been calculated. They incorporate the impact of Magnetic parameter $M$, non-linear stretching parameter $m$, Brownian-motion parameter $Nb$ and thermophoresis parameter $Nt$. Liner as well as quadratic estimations are computed and presented. The signs (plus or minus) of the regression coefficients are used to interpret the direction of relationship between variables. A positive value of coefficient indicates that the dependent variable will have a positive correlation with the independent variable and vice versa. \\
The estimated values of $Nur$ and $Shr$ have been calculated for 625 sets of values of $Nb$, $Nt$ in the set $\lbrace 0.1,0.2,0.3,0.4,0.5 \rbrace$ each, $M$ in set $ \lbrace 0,1,2,5,10 \rbrace$ and $m$ in set $\lbrace 1,2,3,4,5 \rbrace$. The regression analysis is executed for different values of $Pr$ ranging $\lbrace 0.7,1.5,2,4,5 \rbrace$ and $Ln$ ranging $\lbrace 5,10,15,20,25 \rbrace$ with default values for $Pr$ and $Ln$ being $2$ and $10$ respectively. The simple linear regression formula is sufficient for most practical purposes but for high accuracy, quadratic regression analysis has also been performed. \\
The regression estimations can be written as:\\
Linear regression estimation:
\begin{equation}
\begin{split}
Nur_{mre}  = Nu + C_{M}M + C_{L}m+ C_{b}Nb+ C_{t}Nt\\
Shr_{mre} = Shr + C_{M}^{'}M+ C_{L}^{'}m+ C_{b}^{'}Nb+ C_{t}^{'}Nt
\end{split}
\end{equation}
Quadratic regression estimation:
\begin{equation}
\begin{split}
Nur_{mre}  = Nu + C_{M}M + C_{L}m+ C_{b}Nb+ C_{t}Nt + C_{MM}M^{2} + C_{LL}m^{2} \\ + C_{bb}Nb^{2}+ C_{tt}Nt^{2} +C_{Mm}Mm + C_{Mt}MNt+ C_{bL}Nbm +
 C_{bt}NbNt \\
 Shr_{mre} = Shr + C_{M}^{'}M + C_{L}^{'}m+ C_{b}^{'}Nb+ C_{t}^{'}Nt+ C_{MM}^{'}M^{2} + C_{LL}^{'}m^{2} \\ + C_{bb}^{'}Nb^{2}+ C_{tt}^{'}Nt^{2} +C_{Mm}^{'}Mm+ C_{Mt}^{'}MNt+ C_{bL}^{'}Nbm+ C_{bt}^{'}NbNt \\
\end{split}
\end{equation}
where $C_{M},C_{L},C_{b},C_{t},C_{MM},C_{LL},C_{bb},C_{tt},C_{Mm},C_{Mt},C_{bL} $ and $C_{bt}$  are coefficient of Nusselt number estimations and $C_{M}^{'},C_{L}^{'},C_{b}^{'},C_{t}^{'},C_{MM}^{'},C_{LL}^{'},C_{bb}^{'},C_{tt}^{'},$ \\ $C_{Mm}^{'},C_{Mt}^{'},C_{bL}^{'} $ and $C_{bt}^{'}$  are coefficient of Sherwood number estimations. \\
Tables 6 and 7 provide the values of coefficients of equation 27 for $Pr$ =2 and $Ln$ = 10 respectively. From tables 6 and 7, it can be observed that coefficients $C_{M},C_{L},C_{b}, C_{t}$  and $C_{M}^{'},C_{L}^{'}$ are negative, whereas $C_{t}^{'}$ and $C_{b}^{'}$ are positive.\\
This implies that Nusselt number ($Nur$) is a decreasing function of magnetic parameter ($M$), non-linear stretching parameter ($m$), Brownian motion parameter ($Nb$) and thermophoresis parameter ($Nt$). Sherwood number ($Shr$)is a decreasing function of magnetic parameter ($M$) and non-linear stretching parameter ($m$), whereas it is an increasing function of Brownian-motion parameter ($Nb$) and thermophoresis parameter ($Nt$). The results of regression analysis verify the results obtained from FEM. Standard error has been calculated in both cases. For better accuracy, quadratic regression analysis has also been performed. (Table 8 and 9) \\
For $Pr=2$ and $Ln=10$, the linear estimates are:\\
\begin{equation}
\begin{split}
Nur_{mre}  = 0.3866  - 0.0145 M  - 0.0017m -0.1572Nb - 0.0807Nt\\
Shr_{mre} = 0.8398 -0.0353M - 0.0038m + 0.0139Nb + 0.0069Nt
\end{split}
\end{equation}
For $Pr=2$ and $Ln=10$, the quadratic estimates are:\\
\begin{equation}
\begin{split}
Nur_{mre}  = 0.4314 -0.0379M -0.0063m - 0.2023Nb - 0.1243Nt + 0.0021M^{2} \\ + 0.0006m^{2}  + 0.0472Nb^{2}+ 0.0171Nt^{2} +0.0004Mm + 0.0046MNt \\+ 0.0010Nbm + 0.0494NbNt \\
 Shr_{mre} =0.9956 -0.1213M -0.1782m + 0.1578Nb+ 0.1682Nt- 0.0003M^{2}\\ + 0.0092m^{2}  -10.4498Nb^{2} +2.0018Nt^{2} +0.0636Mm +0.0444MNt\\+5.1407Nbm+0.1804NbNt \\
\end{split}
\end{equation}

\section{Code validation}
The numerical procedure used in the problem is validated by using a test case. The mathematical value of magnetic parameter is considered to be zero. The temperature, the nanoparticle concentration and solute concentration are assumed to have a constant value at the boundary. The boundary conditions considered at $y=0$ are as follows:
\begin{equation}
u_{w}=ax^{m}, \;  v=0, \;  T = T_{w}, \; C = C_{w}, \;  \phi = \phi_{w}
\end{equation}
Reduced Nusselt number $Nur$ is calculated and the results obtained from present code are found to be in excellent agreement with the results of Goyal and Bhargava \cite{Ref8} (Table 2).

\section{Results and Discussions}
To study the behaviour of the system, numeric computations has been conducted by varying the values of different controlling parameters that describe flow characteristics. The results are presented both in tabular form and graphically. \\
The profiles of all the functions - velocity ($s^{'}(\eta)$), temperature ($\theta(\eta)$), nanoparticle concentration ($f(\eta)$) and solutal concentration ($\gamma(\eta)$)decrease monotonically with an increase in $\eta$. As $ \eta \rightarrow \infty$, the value of the functions approaches to zero asymptotically. Figure 1 presents the behaviour of nanofluid velocity for magnetic parameter $M$. An intensification in the value of $M$ causes the value of nanofluid velocity to decline. This is because an increase in $M$ leads to an increase in Lorentz drag force which resists the motion of nanofluid. The magnetic parameter can thus, be used to control the velocity of nanofluid flow over the sheet. Figure 2, 3 and 4 illustrates the effect of magnetic parameter $M$ on temperature profile $\theta ( \eta)$, solutal concentration $ \phi ( \eta)$ and nano particle concentration $f( \eta) $ through the boundary layer regime. With an intensification in the value of $M$, the de-accelerating Lorentz drag force intensifies which in turn increases the value of resistance provided to the fluid flow, causing an increase for the profiles of temperature, solutal and nanoparticle concentration.\\
The numerical values of reduced Nusselt number $Nur$, reduced local Sherwood number $Shr$ and reduced nanofluid Sherwood number $Shrn$ are presented in tables 3,4 and 5 respectively for various values of Prandtl number $Pr$, Lewis number $Ln$ and magentic parameter $M$. The values of other parameters are fixed - $ Nb=0.5$, $Nt=0.5$, $Nd=0.2$, $Ld = 0.1$ and $Le = 2.0$. The value of heat transfer rate, regular mass transfer rate and nano mass transfer rate drops down with an intensification in value of magnetic parameter $M$. The rate of heat transfer is calculated by $- \theta^{'} (0)$. An increase in temperature will cause the difference in subsequent temperatures to decrease, thus decreasing the  value of $Nur$. Similar argument can be provided for $Shr$ and $Shrn$. An amplification in values of Prandtl number $Pr$ causes a drop in heat transfer rate and a growth in mass transfer rates. An increase in Lewis number also depicts the same trend as Prandtl number. Lewis number is the fraction of thermal diffusivity over mass diffusivity. An increase in Lewis number causes weakening of mass diffusivity and strengthening of thermal diffusivity. With the lowering of mass diffusivity, the concentration of both solute and nanoparticles increase on the boundary layer thereby increasing the value of $Shr$ and $Shrn$. Similar argument can be provided for $Nur$.\\
The random motion of particles inside a medium is termed as Brownian motion. As name suggests, the value of Brownian motion parameter, $Nb$ describes the intensity of randomness of particles in medium.
Fig. 5a shows the effect of arbitrary motion of particles along with the magnetic parameter on heat transfer rate $Nur$. With an increase in $Nb$, the thermal boundary layer thickens as motion of nanoparticles intensifies. An increase in $M$ leads to increase in Lorentz drag force opposing the motion of the fluid. The combined influence of increase of both the parameters is that the temperature increases, thus reducing the effective heat transfer. Fig. 5b and Fig. 5c shows the effect of Brownian motion along with the magnetic parameter on regular and nano mass transfer rate respectively. An increase in $Nb$ causes an augmentation in arbitrary movement of particles. This causes the warming of the boundary layer which effectively causes the nanoparticles to move away from the wall of the sheet inside the inactive fluid. This increases the deposition of the particles away from the sheet, thereby justifying for the reduced concentration magnitudes ($\phi ( \eta)$ and $f( \eta) $) and an increase in the value of regular mass transfer rate $Shr$ and nano mass transfer rate $Shrn$. \\
The dissemination of particles as a consequence of temperature gradient is termed as thermophoresis. The thermophoresis parameter $Nt$ is used to gauge this effect. Fig. 6a shows the effect of thermophoresis parameter along with the magnetic parameter on heat transfer rate $Nur$. The temperature gradient generates a thermophoretic force, causing the flow direction to move away from the stretching surface. Thus, more and more fluid is heated away from the surface and hence, an increase in $Nt$ results in increasing the temperature within the boundary layer. An increase in magnetic parameter $M$ causes an increase in Lorentz drag force opposing the fluid motion. The simultaneous increase of both $Nt$ and $M$ causes the heat transfer to increase and as a consequence the value of heat  transfer rate $Nur$ falls down. The thermophoretic effect $Nt$ on regular and nano mass transfer rate can be observed in Fig. 6b and 6c. It is quite evident that an escalation in thermophoretic effect $Nt$ will correspond to increased mass flux (because of temperature gradient) causing an appreciable increase in the concentration. A decrease in magnetic parameter $M$ supports the fluid flow motion because of the decrease in Lorentz drag force, causing an increase in mass flux. \\
Fig. 7a, 7b and 7c shows the effect of magnetic parameter $M$ on the reduced Nusselt number $Nur$, reduced local Sherwood number $Shr$ and reduced nanoparticle Sherwood number $Shrn$ as the nonlinear stretching parameter $m$ increases. As the value of magnetic parameter $M$ increases due to intensification of magnetic fiend, the heat transfer rate and regular and nanomass transfer rate decreases. For $M=0$, there is a sudden decrease in the value of the $Nur$, $Shr$ and $Shrn$ for $m<0.5$. For other values of $m$, the decrease is very gradual. The velocity of the stretching of the sheet has negligible effect on heat and mass transfer rates. \\
By controlling the values of parameters discussed above and choosing an appropriate nanofluid depending on application cooling in the process of extrusion can be affectively achieved. The final thickness of the product and the finish of the product can be enhanced with the knowledge of behavior of concentration profiles along the boundary layer.

\section{Conclusion}
The present paper provides an analysis and explanation of behaviour of boundary layer flow of nanofluid atop a stretching sheet moving with non-linear velocity. Two component model of nanofluid incorporating the effects of thermophoresis, diffusiophoresis and Brownian motion along with application of external variable magnetic field is considered for the study. Finite Element Method is employed to numerically compute the results. Regression analysis of the variables is done to establish and verify the relationship among parameters and rates of transfer.
The results concluded from the study are as follows:\\
1.  Amplifying the value of the magnetic parameter $M$ contracts the momentum boundary layer and expands the thermal, solutal and nano-mass volume fraction boundary layer. An external magnetic fluid gives rise to magnetic body force (Lorentz drag force) retarding the fluid motion. By adjusting the external magnetic field, the heat transfer can be controlled. Widespread growth in the field of 'smart' cooling devices is based on this idea.  \\
2. Strengthening the values of the Brownian motion parameter $Nb$ and thermophoresis parameter $Nt$, the local heat transfer rate and local solutal and nano mass concentration reduces for a rise in the value of magnetic parameter $M$. As different combinations of nanoparticles and base fluids have different values for these parameters (i.e. $Nb$ and $Nt$), thus having different heat and mass transfer rates. This idea can be used in customizing both the heat and mass transfer rates for various industrial processes involving the stretching sheet (Extrusion of metal sheets, manufacture of tetrapacks etc.).  \\
3. With an increase in nanofluid Lewis number $Ln$, heat transfer rate decreases while the regular and nano fluid and mass transfer rate increases. $Ln$ is used to distinguish fluid flow when heat and mass transfer happens simultaneously. With an increase in $Ln$, the diffusion because of Brownian motion decreases, hence forcing the concentration to reduce. \\
4. The magnitude of heat, regular and nano mass transfer rates declines with an increase in the magnetic parameter $M$ because of intensified Lorentz drag force. $\vspace{1 cm}$

\begin{flushleft}
Acknowledgements
\end{flushleft}
The first author acknowledges the support of Department of Science and Technology, Government of India, for providing the financial assistance to carry out the present research work.

\begin{flushleft}
Declaration
\end{flushleft}
The authors declare that they have no conflict of interest.\\
\includegraphics[width=.50\columnwidth]{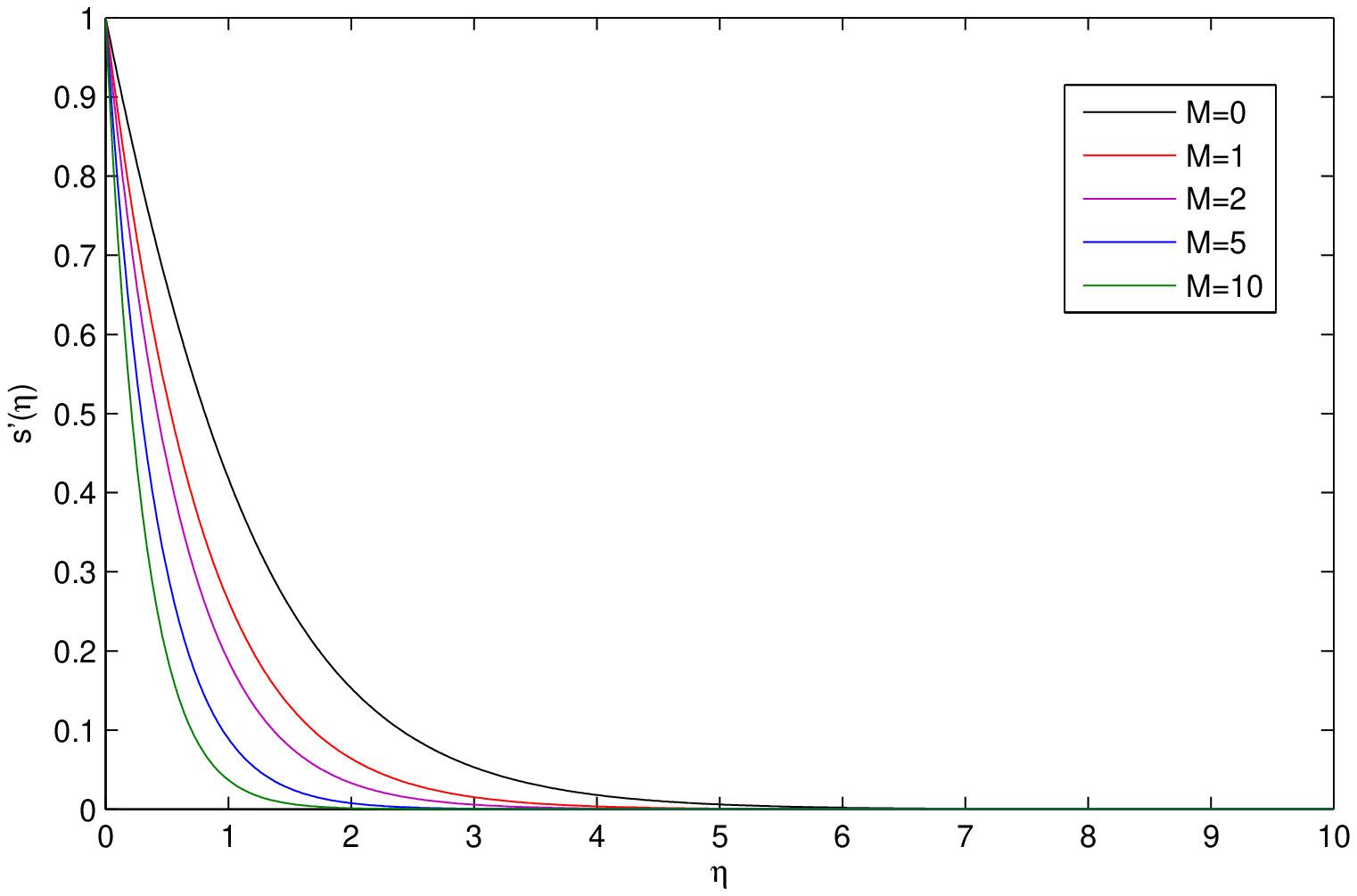}\\
\textbf{Fig. 1 } Effect for magnetic parameter $M$ on velocity distribution.

\includegraphics[width=.50\columnwidth]{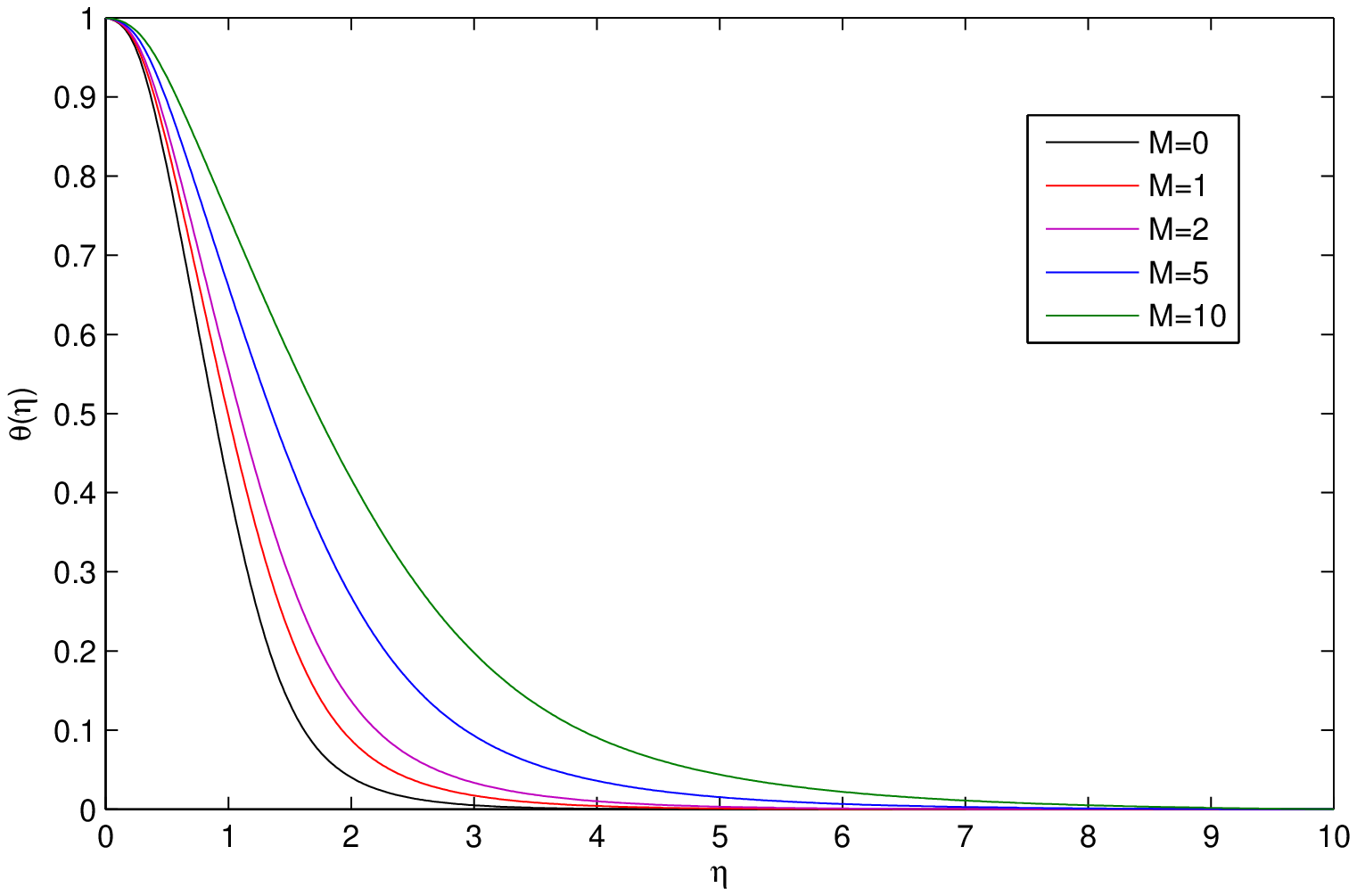}\\
\textbf{Fig. 2} Effect for magnetic parameter $M$ on temperature distribution.

\includegraphics[width=.50\columnwidth]{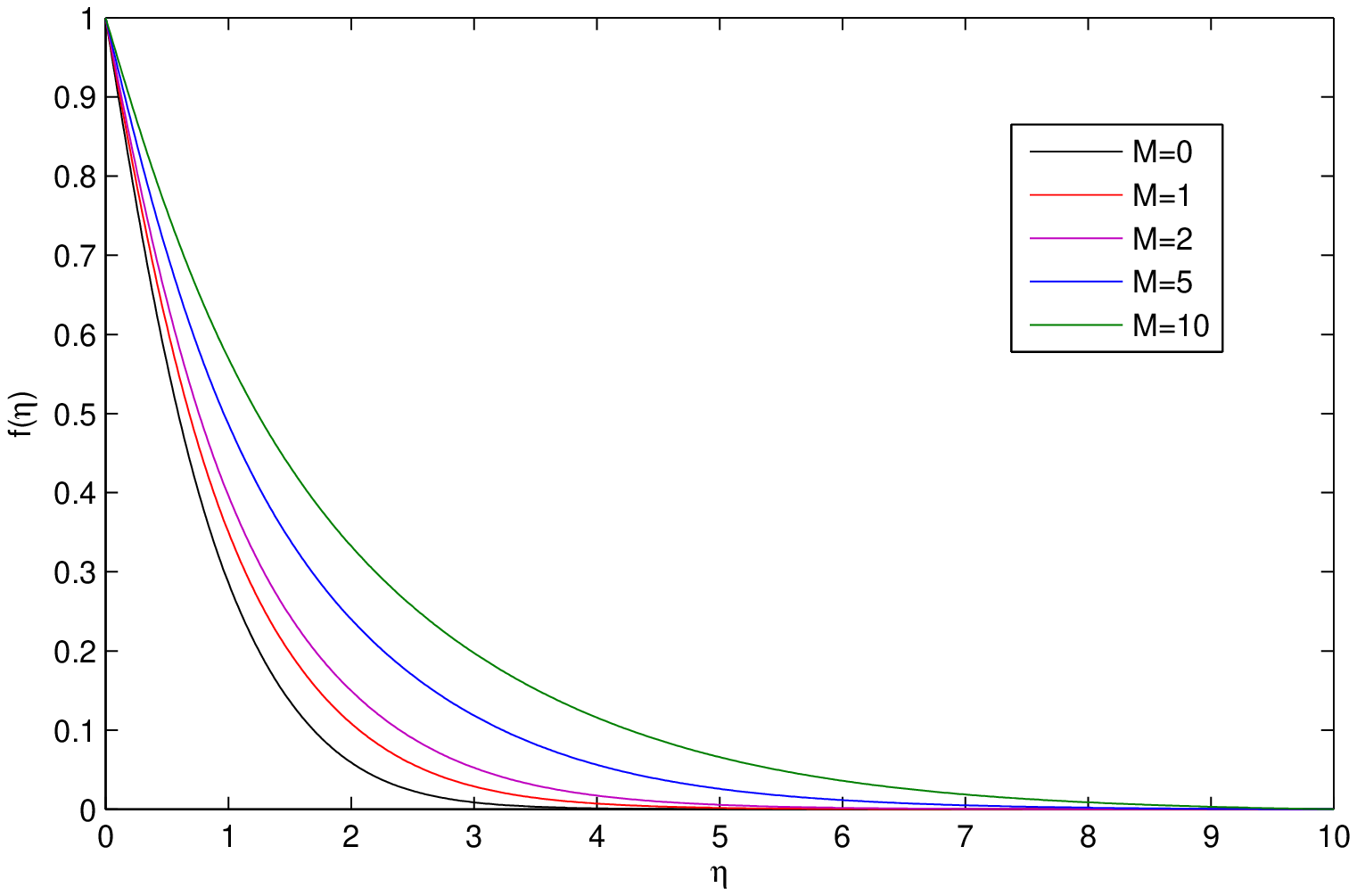}\\
\textbf{Fig. 3 } Effect for magnetic parameter $M$ on nanoparticle concentration.\\

\includegraphics[width=.50\columnwidth]{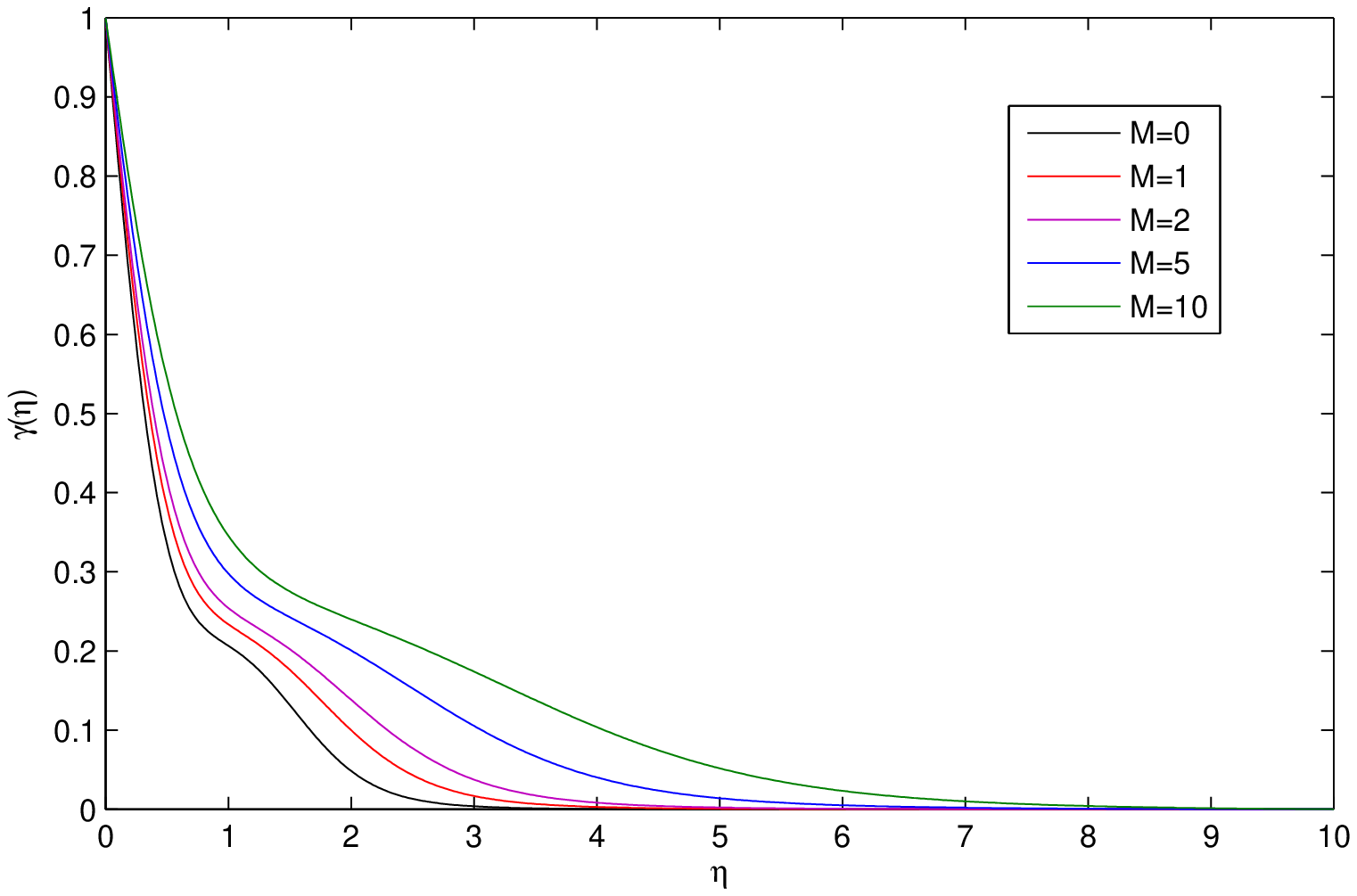}\\
\textbf{Fig. 4} Effect for magnetic parameter $M$ on solutal concentration.\\

\includegraphics[width=.50\columnwidth]{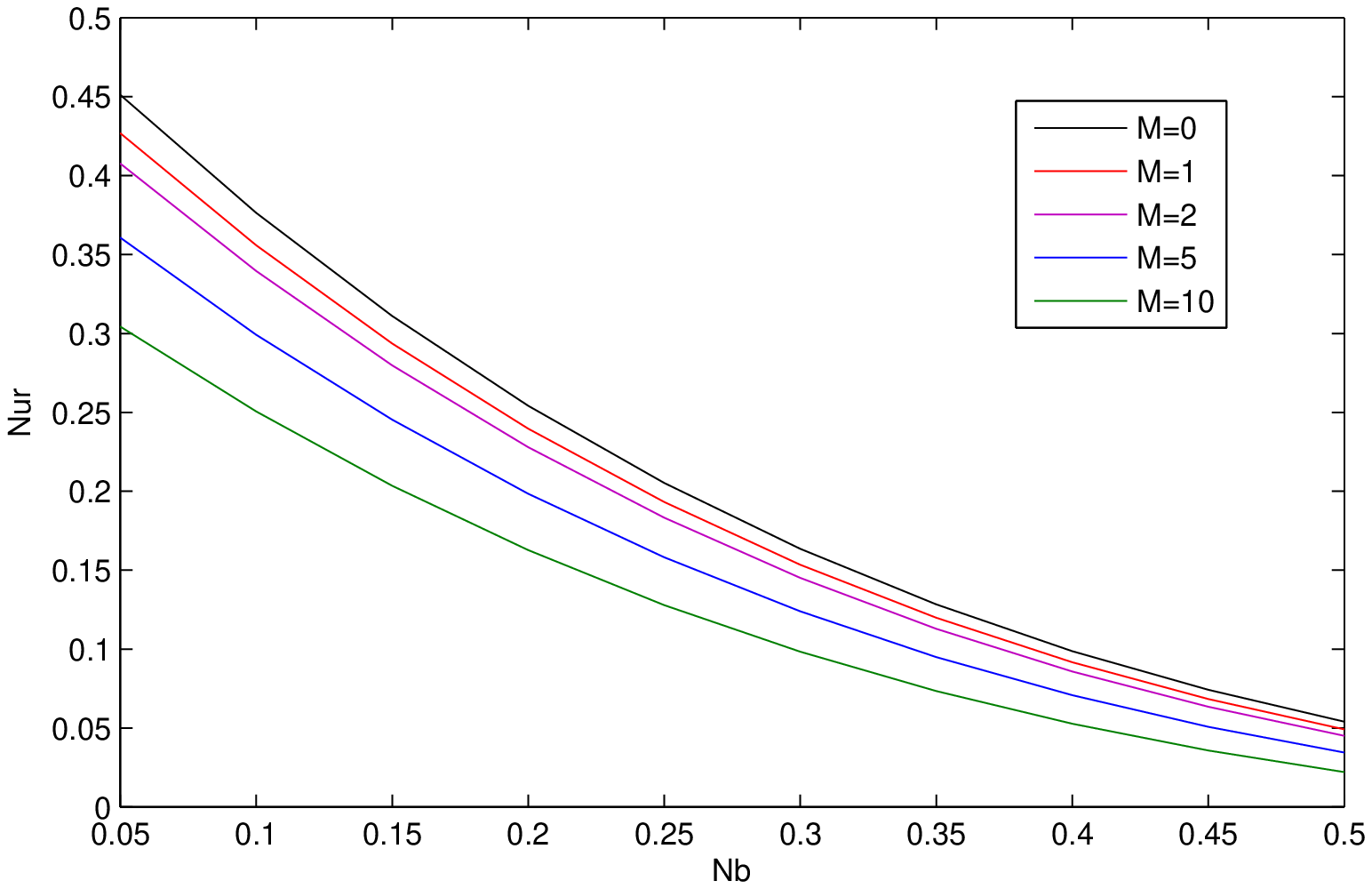}\\
\textbf{Fig. 5a} Effect of Brownian motion parameter $Nb$ and magnetic parameter $M $ on $Nur$.\\

\includegraphics[width=.50\columnwidth]{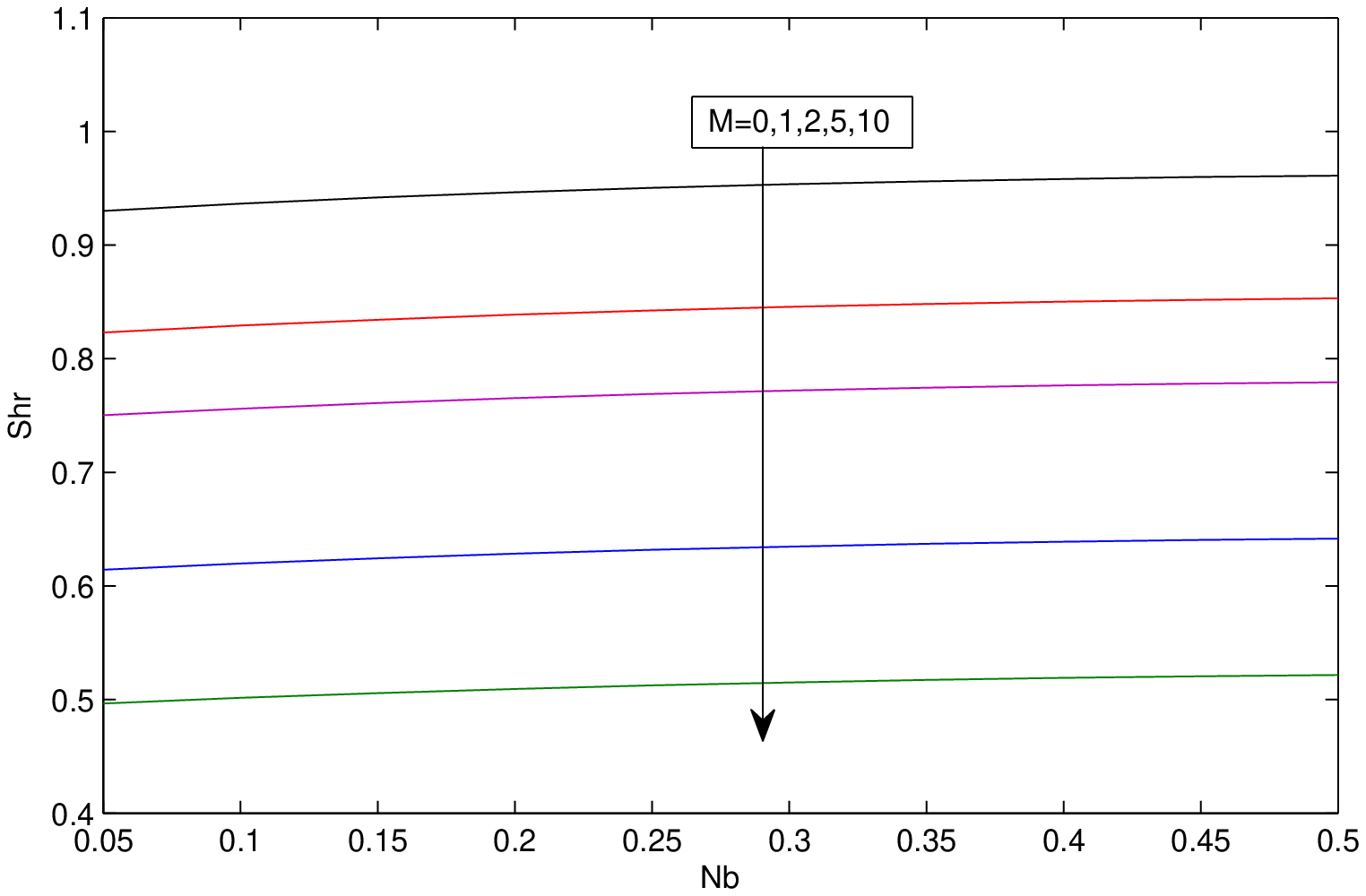}\\
\textbf{Fig. 5b } Effect of Brownian motion parameter $Nb$ and magnetic parameter $M $ on $Shr$.\\

\includegraphics[width=.50\columnwidth]{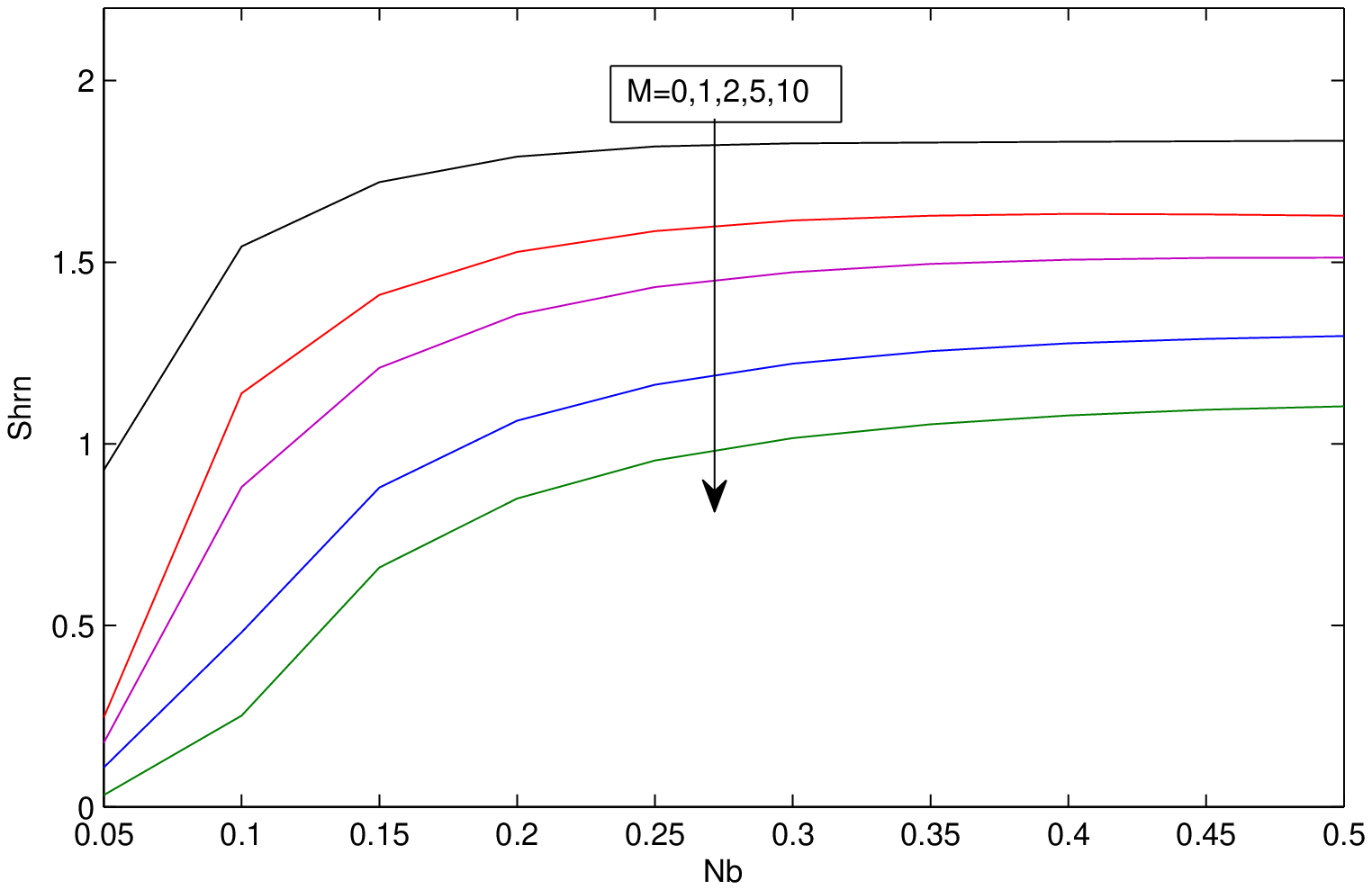}\\
\textbf{Fig. 5c} Effect of Brownian motion parameter $Nb$ and magnetic parameter $M $ on $Shrn$ .\\

\includegraphics[width=.50\columnwidth]{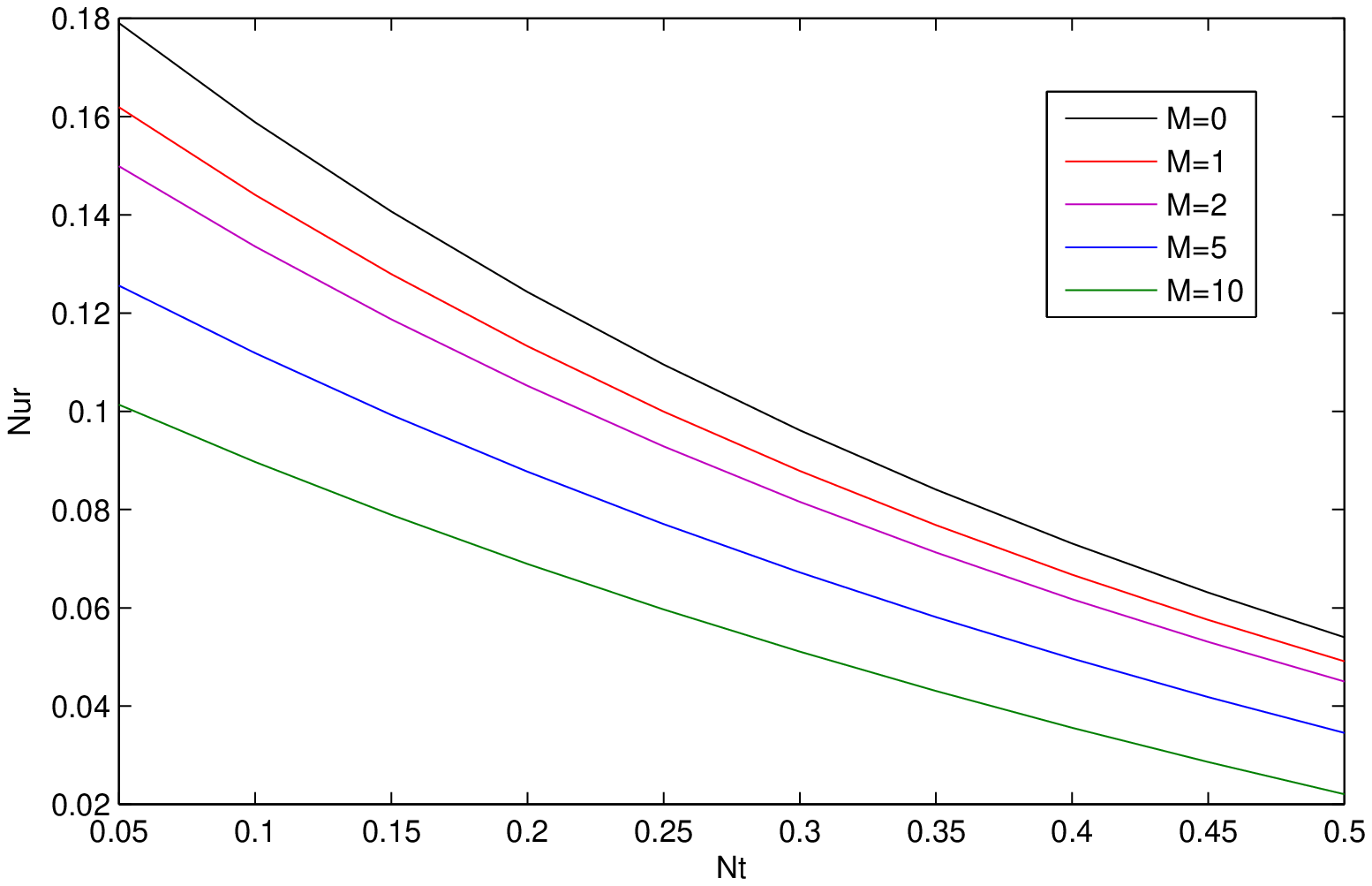}\\
\textbf{Fig. 6a} Effect of thermophoresis parameter $Nt$ and magnetic parameter $M $ on $Nur$.\\

\includegraphics[width=.50\columnwidth]{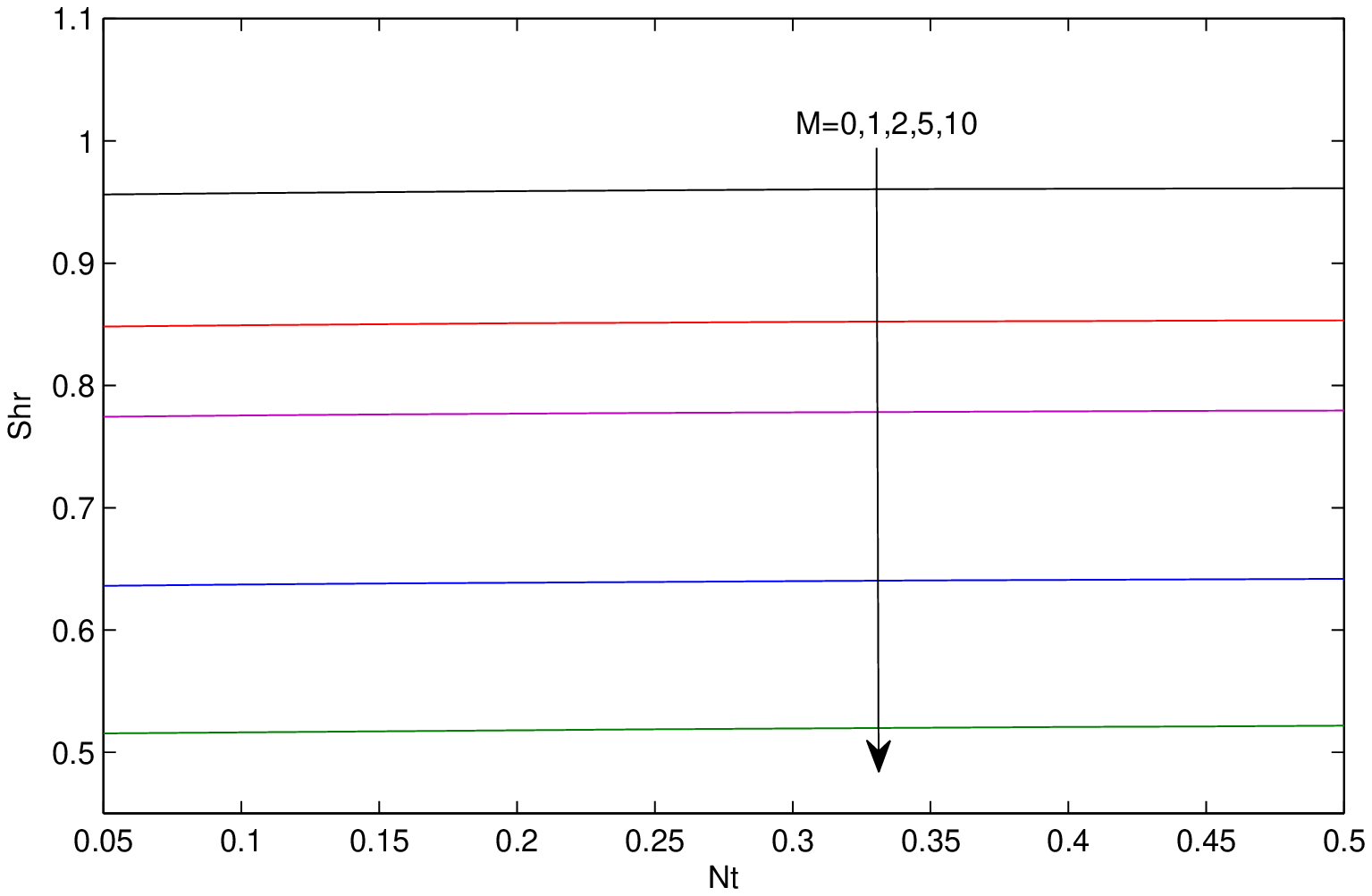}\\
\textbf{Fig. 6b} Effect of thermophoresis parameter $Nt$ and magnetic parameter $M $ on $Shr$.\\

\includegraphics[width=.50\columnwidth]{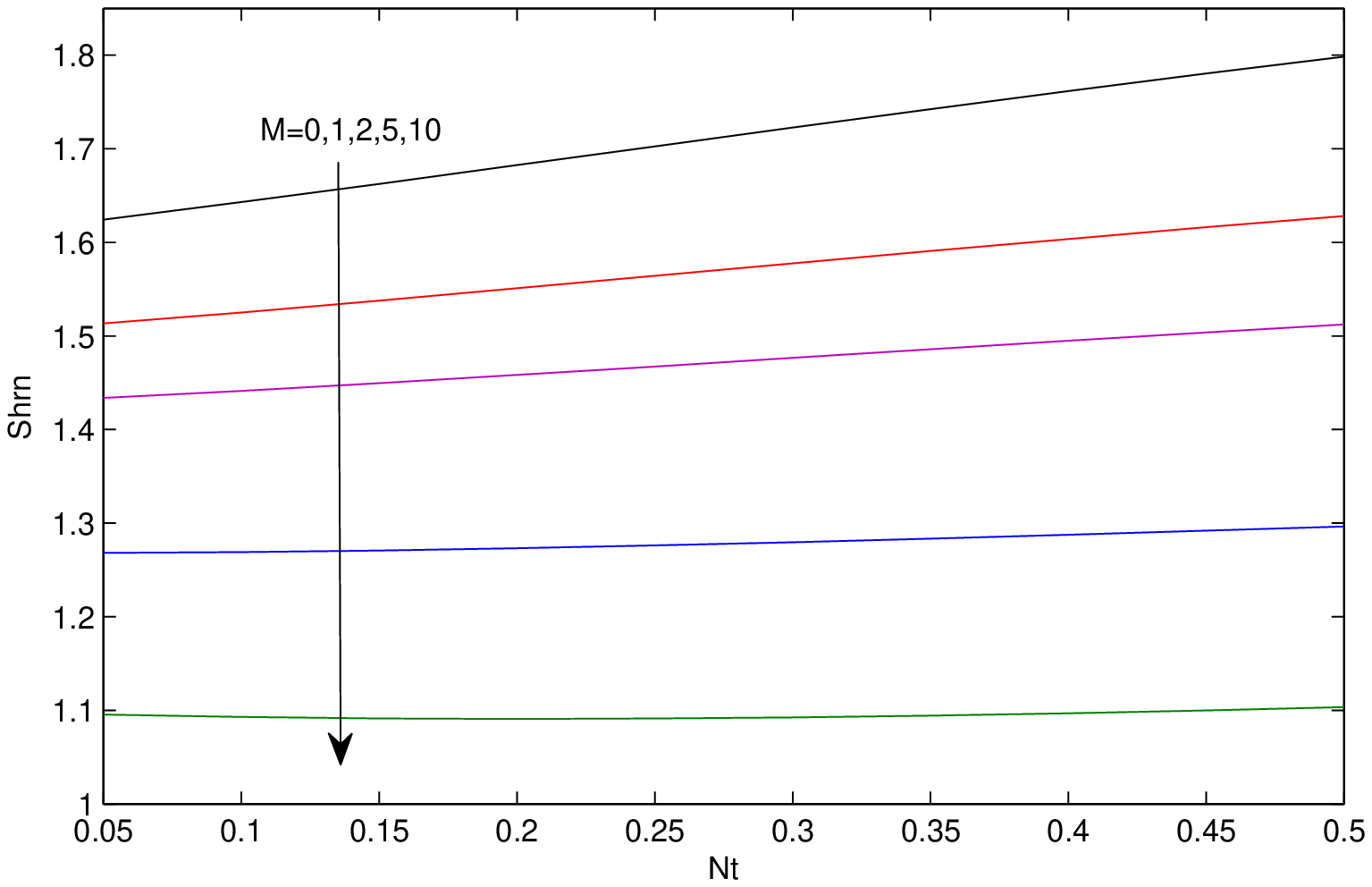}\\
\textbf{Fig. 6c} Effect of thermophoresis parameter $Nt$ and magnetic parameter $M $ on $Shrn$ .\\

\includegraphics[width=.50\columnwidth]{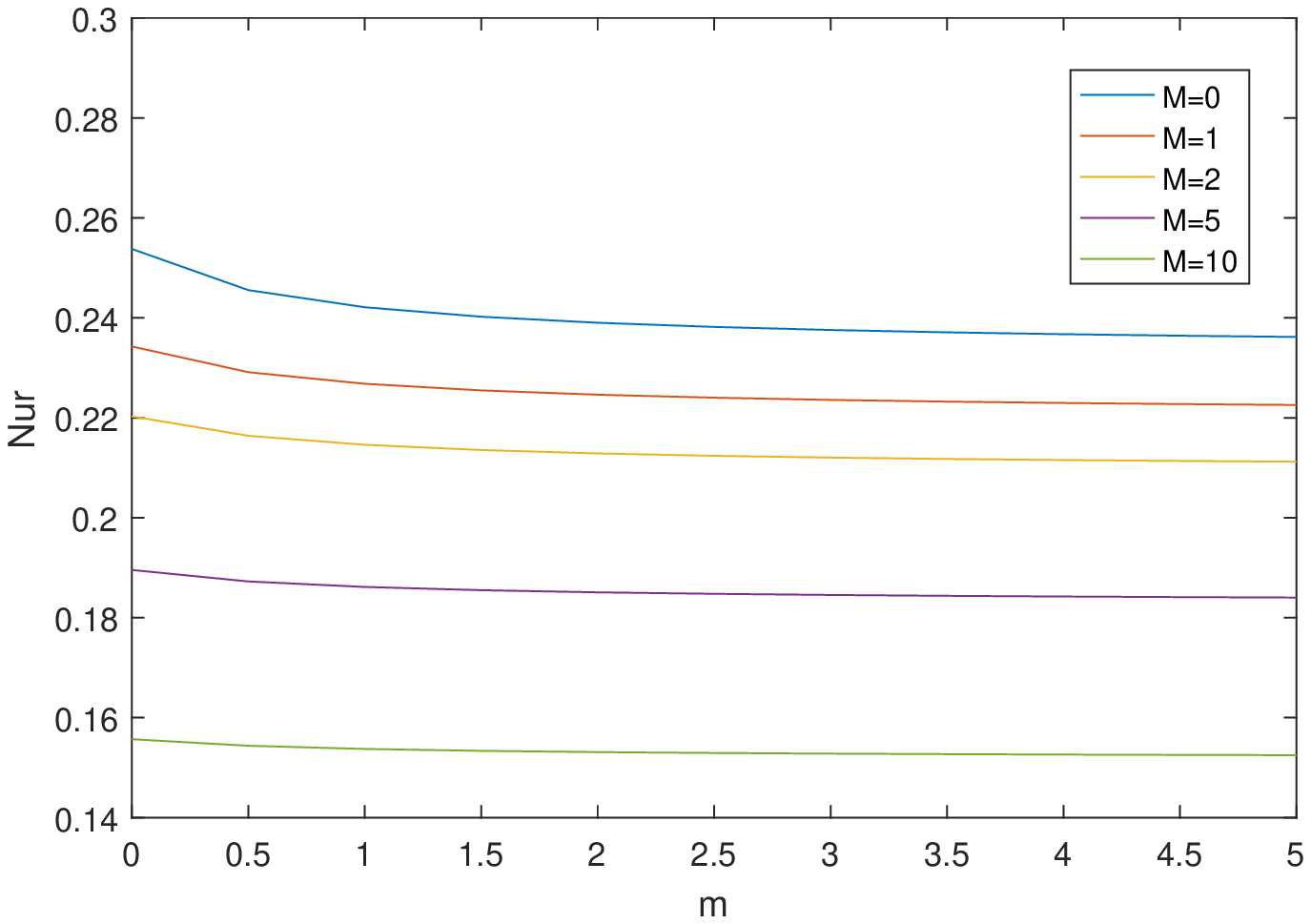}\\
\textbf{Fig. 7a} Effect of stretching parameter $m$ and magnetic parameter $M $ on $Nur$.\\

\includegraphics[width=.50\columnwidth]{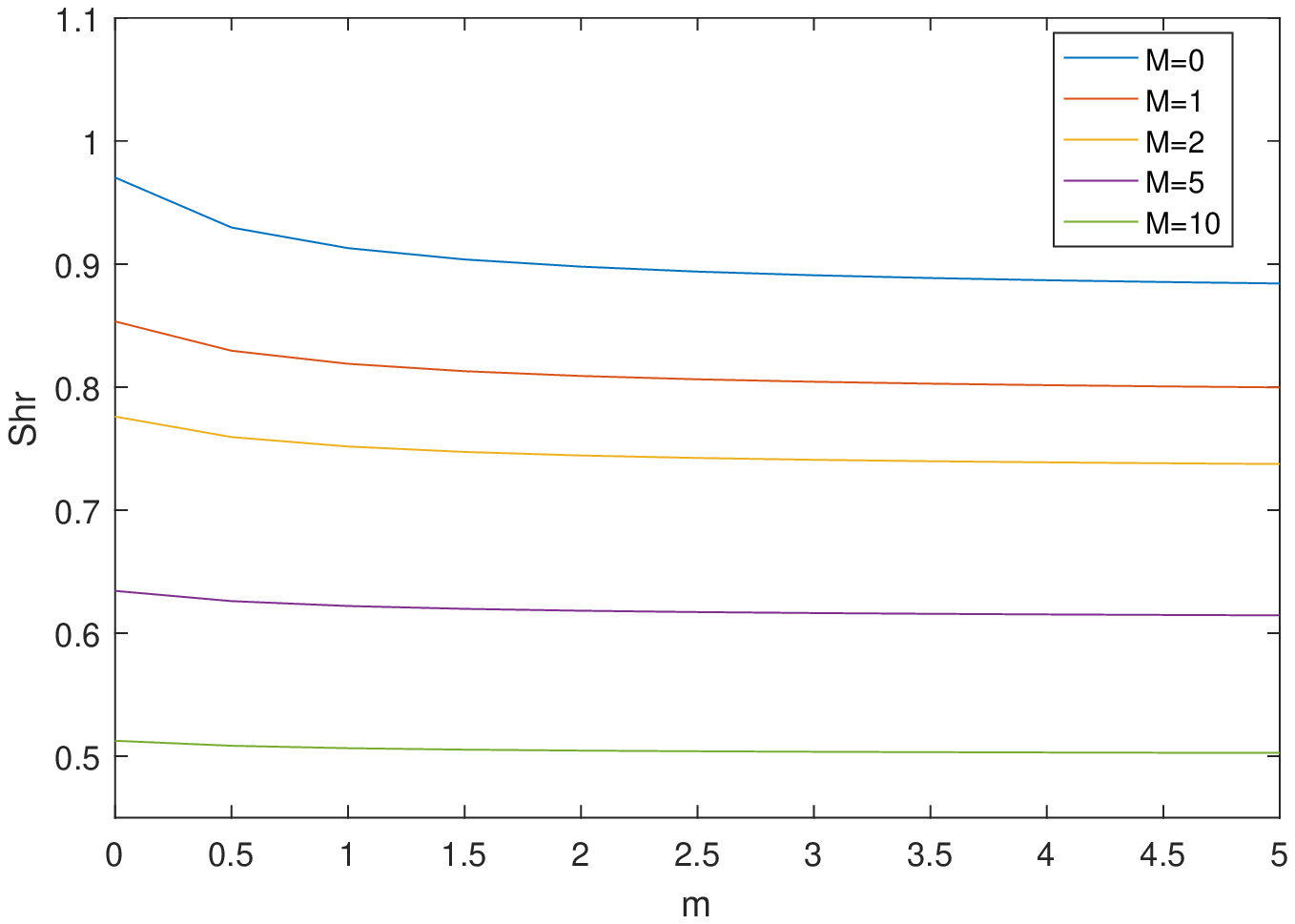}\\
\textbf{Fig. 7b} Effect of stretching parameter $m$ and magnetic parameter $M $ on $Shr$.\\

\includegraphics[width=.50\columnwidth]{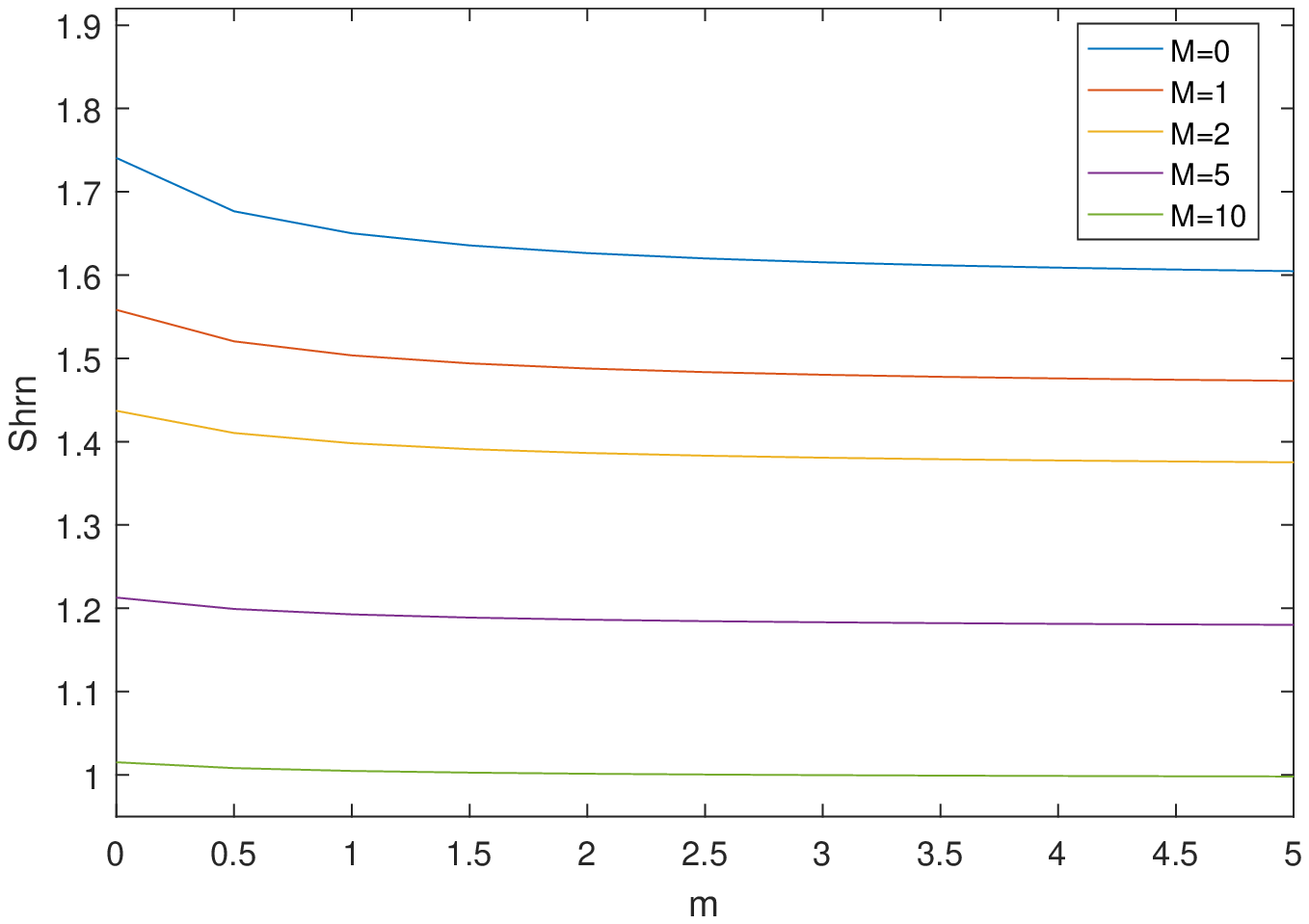}\\
\textbf{Fig. 7c} Effect of stretching parameter $m$ and magnetic parameter $M $ on $Shrn$.\\
\onecolumn
\begin{table}
\begin{tabular}{c | c c c | c c c | c c c}
\hline
Step size &  & $|- \theta '(0)|$    & • &  &$ |- \gamma' (0)|$ & • & • & $| - f'(0)|$ & • \\
\hline
• & $\eta_{\infty}$=8 & $\eta_{\infty}$=10 & $\eta_{\infty}$=12 & $\eta_{\infty}$=8 & $\eta_{\infty}$=10 &$\eta_{\infty}$=12 & $\eta_{\infty}$=8 & $\eta_{\infty}$=10 & $\eta_{\infty}$=12 \\
\hline
0.2 & 0.22340 & 0.22355 & 0.22358 & 0.79974 & 0.79961 & 0.79968 & 0.71123 & 0.71007 & 0.71014 \\
0.1 & 0.21790 & 0.21806 & 0.21812 & 0.80318 & 0.80305 & 0.80309 & 0.71945 & 0.71829 & 0.71829 \\
0.04 & 0.21469 & 0.21485 & 0.21489 & 0.80433 & 0.80420 & 0.80425 & 0.72344 & 0.72228 & 0.72229 \\
0.02 & 0.21364 & 0.21379 & 0.21385 & 0.80456 & 0.80443 & 0.80448 & 0.72461 & 0.72345 & 0.72349 \\
0.01 & 0.21311 & 0.21327 & 0.21332 & 0.80464 & 0.80451 & 0.80457 & 0.72516 & 0.72400 & 0.72406 \\
0.005 & 0.21285 & 0.21301 & 0.21308 & 0.80467 & 0.80454 & 0.80458 & 0.72543 & 0.72426 & 0.72429 \\
0.002 & 0.21178 & 0.21285 & 0.21285 & 0.80498 & 0.80467 & 0.80456 & 0.72597 & 0.72442 & 0.72442 \\
\hline
\end{tabular}
\caption{Calculation of $Nur$, $Shr$ and $Shrn$ when $Nb=Nt=0.5$, $Nd=0.2$, $Ld=0.1$, $Ln=Le=Pr=2.0$, $m=2$, $M=1$.}
\label{table 6:}
\end{table}

\begin{table}
\begin{tabular}{|c|c|c|c|}
\hline
Pr & Ln & Goyal and Bhargava (2014) & Present results \\
\hline
0.7 & 5 & 0.23491 & 0.23492 \\
\hline
• & 15 & 0.22898 & 0.22901 \\
\hline
• & 25 & 0.22722 & 0.22748 \\
\hline
2 & 5 & 0.20980 & 0.20985 \\
\hline
• & 15 & 0.18368 & 0.18376 \\
\hline
• & 25 & 0.177752 & 0.17762 \\
\hline
5 & 5 & 0.05134 & 0.05137 \\
\hline
• & 15 & 0.03259 & 0.03262 \\
\hline
• & 25 & 0.03074 & 0.03078 \\
\hline
\end{tabular}
\caption{Comparison of results for reduced Nusselt number $| - \theta'(0)|$ when $Nb=Nt=0.5$, $Nd=0.2$, $Ld=0.1$, $Le=2.0$, $m=0.2$, $M=0$.}
\label{table 6:}
\end{table}

\begin{table}[h!]
\centering
\begin{tabular}{|c  c  c  c  c  c  c|}
 \hline
 Pr & Ln &  & & Nur & &  \\ [0.7ex]
\hline
  &  & M = 0 & M = 1  & M = 2 & M = 5 & M = 10\\
  \hline
 0.7 & 5 &  0.23493	& 0.18253	& 0.16069 & 0.11731 & 0.09680 \\
 & 15 &  0.22901 & 0.17868 & 0.15759	& 0.11567 & 0.09556\\
  & 25 &  0.22748 & 0.17776 &	0.15688 & 0.11536 & 0.09537
\\
2.0 & 5 &  0.20985 & 0.17717	 & 0.15513 & 0.11600 & 0.08554
 \\
 & 15 & 0.18376 & 0.15559 & 0.13680 & 0.10380 & 0.07818 \\
  & 25 & 0.17762 & 0.15085 & 0.13295 & 0.10156 & 0.07708\\
5.0 & 5 & 0.05137 & 0.04696 & 0.04313 & 0.03330 & 0.02137\\
 & 15 & 0.03262 & 0.02927 & 0.02673 & 0.02134 & 0.01553\\
 & 25 & 0.03078 & 0.02783 & 0.02559 & 0.02102 & 0.01619\\
 \hline
\end{tabular}
\caption{Variations in $Nur$ with $Pr, Ln, M$ when  $m=0.2$, $Nb=Nt=0.5$, $Nd=0.2$, $Ld=0.1$, $ Le=2.0 $}
\label{table:1}
\end{table}

\begin{table}[h!]
\centering
\begin{tabular}{|c  c  c  c  c  c  c|}
 \hline
 Pr & Ln &  & & Shr & &  \\ [0.7ex]
\hline
  &  & M = 0 & M = 1  & M = 2 & M = 5 & M = 10\\
  \hline
 0.7 & 5 &  0.93820	& 0.83680	& 0.76778 & 	0.63807	& 0.52363\\
 & 15 &  0.93926 &  	0.83754	 &  0.76833 &  	0.63833 &  	0.52372\\
  & 25 &  0.93962 &  	0.83780	 &  0.76852 &  	0.63843	 &  0.52377

\\
2.0 & 5 &  0.94554 &  	0.84029	 &  0.76881 &  	0.63652	 &  0.52276\\
 & 15 & 0.94925	 &  0.84340 &  	0.77144	 &  0.63808	 &  0.52335
\\
  & 25 & 0.95045 &  	0.84439 &  	0.77226	 &  0.63856	 &  0.52355
\\
5.0 & 5 & 0.96162 &  	0.85353	 &  0.77991 &  	0.64343 &  	0.52671
\\
 & 15 & 0.96518	 &  0.85681	 &  0.78287	 &  0.64540 &  	0.52736
\\
 & 25 & 0.96643	 &  0.85792	 &  0.78385	 &  0.64602	 &  0.52757
\\
 \hline
\end{tabular}
\caption{Variations in $Shr$ with $Pr, Ln, M$ when  $m=0.2$, $Nb=Nt=0.5$, $Nd=0.2$, $Ld=0.1$, $ Le=2.0 $}
\label{table:2}
\end{table}

\begin{table}[h!]
\centering
\begin{tabular}{|c  c  c  c  c  c  c|}
 \hline
 Pr & Ln &  & & Shrn & &  \\ [0.7ex]
\hline
  &  & M = 0 & M = 1  & M = 2 & M = 5 & M = 10\\
  \hline
 0.7 & 5 &  1.57195	 & 1.45777	 & 1.38186 & 1.23351	& 1.08238\\
 & 15 &  2.92396 & 	2.80605 & 	2.72508	 & 2.55830	 & 2.37143
\\
  & 25 &  3.83781	 & 3.71971	 & 3.63819 & 	3.46903 & 	3.27686

\\
2.0 & 5 &  1.65367 & 	1.50707 & 	1.40920 & 	1.22965	 & 1.07116\\
 & 15 & 3.03136	 & 2.88514 & 	2.78399	 & 2.58319 & 	2.37689

\\
  & 25 &3.95140 & 	3.80694 & 	3.70638	 & 3.50391	 & 3.29039

\\
5.0 & 5 & 1.80158	 & 1.63100 & 	1.51497 & 	1.29953 & 	1.11024

\\
 & 15 & 3.15331	 & 2.99210	 & 2.87875	 & 2.65060 & 	2.41733

\\
 & 25 & 4.06256	& 3.90665 & 3.79633	& 3.57039	& 3.33174

\\
 \hline
\end{tabular}
\caption{Variations in $Shrn$ with $Pr, Ln, M$ when $m=0.2$, $Nb=Nt=0.5$, $Nd=0.2$, $Ld=0.1$, $ Le=2.0 $}
\label{table:3}
\end{table}

\begin{table}
\begin{tabular}{ccccccc}
\hline
Pr & Nur & $C_{M}$ & $C_{L}$ & $C_{b}$ & $C_{t}$ & Standard error \\
\hline
0.7 & 0.6731 & -0.0163 & -0.0015 & -0.1943 & -0.0934 & 0.0016 \\
\hline
1.5 & 0.5281 & -0.0331 & -0.0017 & -0.1760 & -0.0884 & 0.0057 \\
\hline
2 & 0.3866 & -0.0145 & -0.0017 & -0.1574 & -0.0807 & 0.0120 \\
\hline
4 & 0.3095 & -0.0201 & -0.0021 & -0.1465 & -0.0769 & 0.0308 \\
\hline
5 & 0.2465 & -0.0159 & -0.0031 & -0.1396 & -0.0684 & 0.0561 \\
\hline
Pr & Shr & $C'_{M}$ & $C'_{L}$ & $C'_{b}$ & $C'_{t}$ & Standard error \\
\hline
0.7 & 0.8103 & -0.0274 & -0.0026 & 0.0117 & 0.0024 & -0.0087 \\
\hline
1.5 & 0.8295 & -0.0309 & -0.0031 & 0.0128 & 0.0054 & 0.0014 \\
\hline
2 & 0.8398 & -0.0353 & -0.0038 & 0.0139 & 0.0069 & 0.0092 \\
\hline
4 & 0.8402 & -0.0412 & -0.0039 & 0.0151 & 0.0087 & 0.0105 \\
\hline
5 & 0.8497 & -0.0461 & -0.0047 & 0.0167 & 0.0096 & 0.0326 \\
\hline
\end{tabular}
\caption{Linear regression coefficients of $Nur_{mre}$ and $Shr_{mre}$ for $Ln$=10}
\label{table:4}
\end{table}

\begin{table}
\begin{tabular}{ccccccc}
\hline
Ln & Nur & $C_{M}$ & $C_{L}$ & $C_{b}$ & $C_{t}$ & Standard error \\
\hline
5 & 0.3872 & -0.0162 & -0.0009 & -0.1585 & -0.0816 & 0.0113 \\
\hline
10 & 0.3866 & -0.0144 & -0.0017 & -0.1573 & -0.0807 & 0.0120 \\
\hline
15 & 0.3846 & -0.0136 & -0.0032 & -0.1552 & -0.0797 & 0.0189 \\
\hline
20 & 0.3838 & -0.0129 & -0.0041 & -0.1546 & -0.0786 & 0.0195 \\
\hline
25 & 0.3829 & -0.0120 & -0.0055 & -0.1538 & -0.0773 & 0.0204 \\
\hline
Ln & Shr & $C'_{M}$ & $C'_{L}$ & $C'_{b}$ & $C'_{t}$ & Standard error \\
\hline
5 & 0.8377 & -0.0348 n& -0.0027 & 0.0126 & 0.0058 & 0.0084 \\
\hline
10 & 0.8398 & -0.0353 & -0.0038 & 0.0139 & 0.0069 & 0.0092 \\
\hline
15 & 0.8412 & -0.0367 & -0.0045 & 0.0152 & 0.0077 & 0.0111 \\
\hline
20 & 0.8425 & -0.0379 & -0.0057 & 0.0167 & 0.0085 & 0.0126 \\
\hline
25 & 0.8438 & -0.0392 & -0.0064 & 0.0183 & 0.0096 & 0.0158 \\
\hline
\end{tabular}
\caption{Linear regression coefficients of $Nur_{mre}$ and $Shr_{mre}$ for $Pr$=2}
\label{table:5}
\end{table}

\begin{table}
\begin{tabular}{cccccc}
\hline
Pr & 0.7 & 1.5 & 2 & 4 & 5 \\
\hline
Nur & 0.2891 & 0.3178 & 0.4314 & 0.5194 & 0.6104\\
\hline
$C_{M}$ & -0.0619 & -0.0259 & -0.0379 & -0.0499 & -0.0139 \\
\hline
$C_{L}$ & -0.0085 & -0.0052 & -0.0063 & -0.0074 & -0.0041 \\
\hline
$C_{b}$ & -0.2069 & -0.2415 & -0.2023 & -0.2046 & -0.1977 \\
\hline
$C_{t}$ & -0.1281 & -0.1224 & -0.1243 & -0.1262 & -0.1205 \\
\hline
$C_{MM}$ & 0.0058 & 0.0032 & 0.0021 & 0.0018 & 0.0007 \\
\hline
$C_{LL}$ & 0.0003 & 0.0007 & 0.0006 & 0.0004 & 0.0008 \\
\hline
$C_{bb}$ & 0.0448 & 0.0484 & 0.0472 & 0.0460 & 0.0496 \\
\hline
$C_{tt}$ & 0.0129 & 0.0192 & 0.0171 & 0.0150 & 0.0213 \\
\hline
$C_{ML}$ & 0.0001 & 0.0005 & 0.0004 & 0.0002 & 0.0007 \\
\hline
$C_{Lb}$ & 0.0028 & 0.0055 & 0.0046 & 0.0037 & 0.0064 \\
\hline
$C_{bt}$ & 0.0008 & 0.0011 & 0.0010 & 0.0009 & 0.0012 \\
\hline
$C_{tM}$ & 0.0468 & 0.0507 & 0.0494 & 0.0481 & 0.0520 \\
\hline
Standard Error & 0.0542 & 0.0652 & 0.0723 & 0.0831 & 0.0984 \\
\hline
\end{tabular}
\caption{Quadratic regression coefficients of $Nur_{mre}$  for $Ln$=10}
\label{table:6}
\end{table}

\begin{table}
\begin{tabular}{cccccc}
\hline
Ln & 5 & 10 & 15 & 20 & 25 \\
\hline
Shr & 0.9813 & 0.9956 & 0.9995 & 1.006 & 1.0128 \\
\hline
$C'_{M}$ & -0.1345 & -0.1213 & -0.1177 & -0.1117 & -0.1053 \\
\hline
$C'_{m}$ & -0.1936 & -0.1782 & -0.1742 & -0.1673 & -0.1589 \\
\hline
$C'_{b}$ & -0.1765 & -0.1578 & -0.1527 & -0.1442 & -0.1351 \\
\hline
$C'_{t}$ & -0.1847 & -0.1682 & -0.1637 & -0.1562 & -0.1482 \\
\hline
$C'_{MM}$ & -0.0013 & -0.0003 & -0.0007 & -0.0002 & -0.0001 \\
\hline
$C'_{mm}$ & 0.0065 & 0.0092 & 0.0097 & 0.0107 & 0.0120 \\
\hline
$C'_{bb}$ & -10.4619 & -10.4498 & -10.4465 & -10.4418 & -9.9988 \\
\hline
$C'_{tt}$ & 1.9853 & 2.0018 & 2.0063 & 2.0318 & 2.0217 \\
\hline
$C'_{ML}$ & 0.0621 & 0.0636 & 0.0642 & 0.0639 & 0.0639 \\
\hline
$C'_{Lb}$ & 0.0246 & 0.0444 & 0.0498 & 0.0588 & 0.0683 \\
\hline
$C'_{bt}$ & 5.1198 & 5.1407 & 5.1464 & 5.1559 & 5.1659 \\
\hline
$C'_{tM}$ & 0.1573 & 0.1804 & 0.1867 & 0.1972 & 0.2083 \\
\hline
Standard Error & 0.1065 & 0.0945 & 0.0885 & 0.0836 & 0.0785 \\
\hline
\end{tabular}
\caption{Quadratic regression coefficients of $Shr_{mre}$  for $Pr$=2}
\label{table:6}
\end{table}

\onecolumn

\end{document}